\documentclass[preprint,12pt]{elsarticle}

\usepackage{graphics}
\usepackage{epsfig}

\usepackage{amssymb}

\begin{document}

\begin{frontmatter}




\title{The neutron long counter NERO for studies of $\beta$-delayed neutron emission in the r-process}

\author[NSCL,JINA1]{J.~Pereira\corref{jorge}}
\ead{pereira@nscl.msu.edu}
\cortext[jorge]{Corresponding author. Tel.: +1 517 9087428; fax: +1 517 3535967}
\author[NSCL,JINA1,PHDEP]{P.~Hosmer \fnref{hosmer}}
\fntext[hosmer]{Present address: Navy Nuclear Power School, Goose Creek, South Carolina, USA}
\author[NSCL,JINA1,PHDEP]{G.~Lorusso}
\author[NSCL,JINA1]{P.~Santi \fnref{santi}}
\fntext[santi]{Present address: Los Alamos National Laboratory, MS E540, Los Alamos, New Mexico, USA}
\author[ISNA,NDAME,JINA2]{A.~Couture}
\author[ISNA,NDAME,JINA2]{J.~Daly}
\author[NSCL,JINA1]{M.~Del~Santo}
\author[NSCL,JINA1,PHDEP]{T.~Elliot}
\author[ISNA,NDAME,JINA2]{J.~G\"orres}
\author[MAINZ,JINA1]{C.~Herlitzius}
\author[MAINZ2,VISTARS]{K.-L.~Kratz}
\author[ISNA,NDAME,JINA2]{L.O.~Lamm}
\author[ISNA,NDAME,JINA2]{H.Y.~Lee}
\author[NSCL,JINA1]{F.~Montes}
\author[NSCL,JINA1,PHDEP]{M.~Ouellette}
\author[NSCL,JINA1,PHDEP]{E.~Pellegrini}
\author[PACIFIC]{P.~Reeder}
\author[NSCL,JINA1,PHDEP]{H.~Schatz}
\author[MAINZ,VISTARS]{F.~Schertz}
\author[NSCL,JINA1,IFK]{L.~Schnorrenberger}
\author[NSCL,JINA1,PHDEP]{K.~Smith}
\author[ISNA,NDAME,JINA2]{E.~Stech}
\author[ISNA,NDAME,JINA2]{E.~Strandberg}
\author[ISNA,NDAME,JINA2]{C.~Ugalde}
\author[ISNA,NDAME,JINA2]{M.~Wiescher}
\author[ISNA,NDAME,JINA2]{A.~W\"{o}hr}
\address[NSCL]{National Superconducting Cyclotron Laboratory, Michigan State University,\\ East Lansing, Michigan, USA}
\address[JINA1]{Joint Institute for Nuclear Astrophysics, Michigan State University,\\ East Lansing, Michigan, USA}
\address[PHDEP]{Department of Physics and Astronomy, Michigan State University,\\ East Lansing, Michigan, USA}
\address[MAINZ]{Institut f\"{u}r Kernchemie, Universit\"{a}t Mainz, Mainz, Germany}
\address[MAINZ2]{Max-Planck-Institut f\"{u}r Chemie, Universit\"{a}t Mainz, Mainz, Germany}
\address[VISTARS]{Virtuelles Institut f\"{u}r Struktur der Kerne and Nuklearer Astrophysik, Mainz, Germany}
\address[IFK]{Institut f\"{u}r Kernphysik, TU Darmstadt, Darmstadt, Germany}
\address[ISNA]{Institute of Structure and Nuclear Astrophysics, University of Notre Dame,\\ South Bend, Indiana, USA}
\address[NDAME]{Department of Physics and Astronomy, University of Notre Dame,\\ South Bend, Indiana, USA}
\address[JINA2]{Joint Institute for Nuclear Astrophysics, University of Notre Dame,\\ South Bend, Indiana, USA}
\address[PACIFIC]{Pacific Northwest National Laboratory, Richland, Washington, USA}





\begin{abstract}
The neutron long counter NERO was built at the National Superconducting Cyclotron Laboratory (NSCL), Michigan State University, for measuring $\beta$-delayed neutron-emission probabilities. The detector was designed to work in conjunction with
a $\beta$-decay implantation station, so that $\beta$ decays and $\beta$-delayed neutrons emitted from implanted nuclei can be measured simultaneously. 
The 
high efficiency of about 40\%, for the range of energies of interest, along with the small background, 
are crucial for 
measuring $\beta$-delayed neutron emission branchings for neutron-rich r-process nuclei produced as low intensity fragmentation beams in in-flight separator facilities.
\end{abstract}


\end{frontmatter}


\section{Introduction}
\label{sec:introduction}
The emission of $\beta$-delayed neutrons by neutron-rich nuclei significantly influences
\cite{Kra93} the nucleosynthesis of heavy elements in the rapid (r-) neutron-capture process~\cite{B2FH,Cam57}. This decay mechanism competes with the $\beta$ decay of r-process nuclei towards the valley of stability
and serves as an additional source of neutrons in late stages of the r-process~\cite{Far06}. Measurements of $\beta$-delayed neutron emission probabilities ($P_{n}$) are needed
for reliable r-process model calculations, and to test the astrophysical assumptions in various r-process models by comparing their final abundance predictions with observations.

From
a nuclear-structure point of view, the $P_{n}$ value provides model constraints at low
beam intensities where $\gamma$-spectroscopy is difficult. The $P_{n}$ value
probes $\beta$-decay strength at excitation
energies slightly above the neutron threshold. It therefore provides nuclear structure
information complementary to $\beta$-decay, which often favors low energy $\beta$-decay
strength owing to the larger phase space (see for example ~\cite{Mon06,Per09}).

The experimental determination of $P_{n}$ requires the measurement of $\beta$-delayed neutrons in coincidence with the $\beta$ particles emitted from the nucleus of interest. This is particularly challenging for nuclei near or at the r-process path due to their very low production rates and the relatively short half-lives---of the order of 10--100 milliseconds. Experiments performed at ISOL-type facilities have successfully exploited the use of neutron long counters (NLC)~\cite{Gro97} to measure $P_{n}$ values of neutron-rich nuclei (see, for instance, the compilations of Refs.~\cite{Rud93,Pfe02}).
NLCs generally consist of a series of gas proportional counters embedded into a moderator block used to thermalize the neutrons prior to their detection. Performance requirements include
a high detection efficiency for neutron energies ranging from a few keV to $\approx$1~MeV.
Because the detector does not measure the energy of individual neutrons, variations of the efficiency as a function of energy have to be minimized as much as possible as they otherwise can translate into uncertainties in the measured $P_{n}$. Our goal was to keep detector induced systematic uncertainties well below the 10\% level. Measurements at that level of accuracy are a dramatic improvement over theoretical predictions, and ensure that other uncertainties dominate astrophysical and nuclear structure models. With systematic errors at that level, statistical errors will tend to dominate in practice, as the
most interesting isotopes will typically be produced at rather low rates.


We report here the development of NERO, a new NLC at National Superconducting Cyclotron
Laboratory (NSCL)
suitable for use with fast radioactive beams produced by in-flight fragmentation. This technique
provides exotic beams without some of the limitations induced by chemistry-based target-extraction techniques. The short time required to transport, separate, and identify the produced fragments, typically less than few hundred ns, makes it possible to study the very short-lived nuclei in the r-process. The fragments of interest are implanted in an active catcher that is part of the NSCL Beta Counting System (BCS). Implantation of a fragment and emitted $\beta$ particles are detected event-by-event. The correlation of decays with a previously implanted nucleus requires large area highly pixelated catchers, typically double-sided silicon strip detectors (DSSDs). The challenge in the design of NERO was to include a large cylindrical cavity capable of accommodating such a system, while still fulfilling the performance requirements for the detection efficiency. The final design was inspired by existing NLC detectors such as the Mainz Neutron Detector~\cite{Meh96}.


\section{Technical aspects}
\label{sec:technical}
\subsection{Design}
\label{sec:design}
\begin{figure}[h!]
\begin{center}
\includegraphics[height=4.5cm]{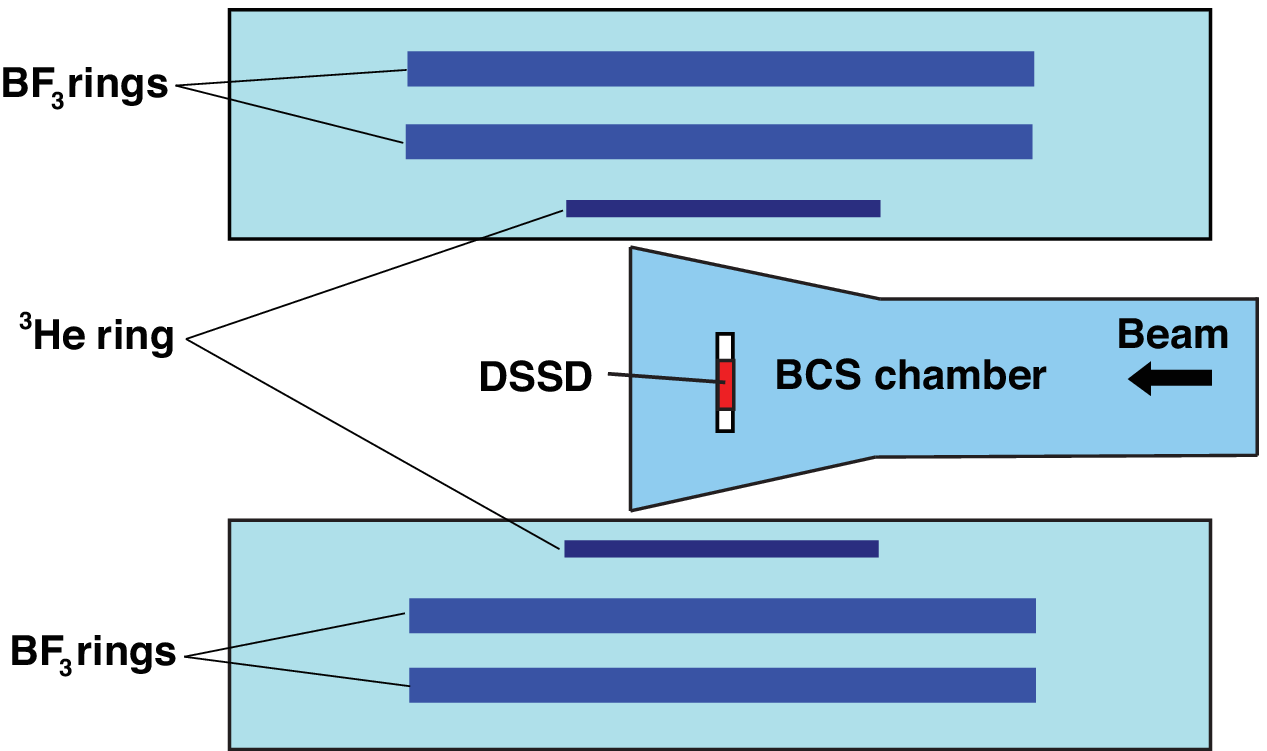} \hfill
\includegraphics[height=4.5cm]{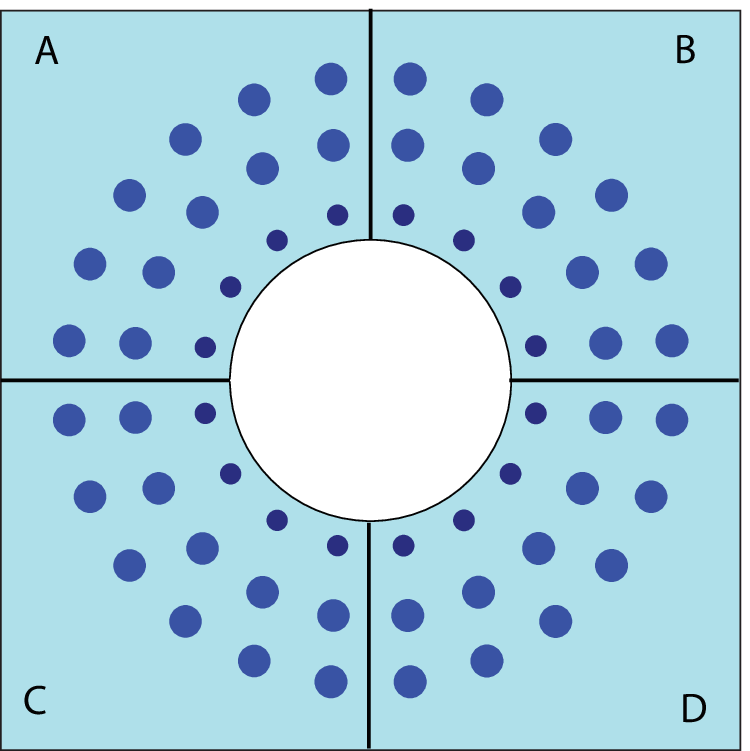}
\caption{Schematic drawings of the NERO detector. Left: Side view showing the BCS chamber located inside of NERO with the DSSD at the central position. Right: backside showing the cylindrical cavity to house the BCS and the three concentric rings of gas-filled proportional counters. The labels A, B, C and D designate the four quadrants.}
\label{fig:NERO}
\end{center}
\end{figure}

The detector system consists of a 60$\times$60$\times$80~cm$^{3}$ polyethylene matrix (density 0.93(1)~g/cm$^{3}$) with its long symmetry axis aligned with the beam. Along the beam axis,
the matrix has a cylindrical cavity with a diameter of 22.8~cm to accommodate the BCS
(see Fig.~\ref{fig:NERO}, left).

NERO includes three different types of cylindrical proportional counters manufactured by Reuter-Stokes: filled with $^{3}$He (models RS-P4-0814-207 and RS-P4-0810-104), and filled with BF$_{3}$ (model RS-P1-1620-205) (see Tab.~\ref{TabCounters} for details). Sixty 
of these detectors are arranged in three concentric rings around the central symmetry axis, allowing for a nearly 4$\pi$ solid angle coverage around the implantation detector (see Fig.~\ref{fig:NERO}, right). The optimum detector configuration was found using the MCNP code~\cite{MCNP} to calculate the neutron-detection efficiency for different geometries, moderating materials, and number and arrangement of various types of proportional counters. Interactions of neutrons with the different detector materials were calculated, using the ENDF/B-VI~\cite{Hen94} cross-sections in the energy range 10$^{-5}$~eV to 20~MeV. The influence of different environments such as laboratory floor and wall configurations were investigated but were found to be negligible. According to these calculations, most of the neutrons emitted from the center of NERO are detected in the innermost ring. Therefore, sixteen of the more compact and efficient  $^{3}$He gas-filled proportional counters are mounted in the innermost ring at  a radius of 13.6~cm. For the middle and outer rings at  radii of 19.2~cm and 24.8~cm we use twenty and twenty-four BF$_{3}$ proportional counters, respectively. The BF$_{3}$ counters are longer allowing one to cover more solid angle, and their efficiency just compensates the decreasing
efficiency of the inner ring with increasing neutron energy.

To facilitate transportation and assembly, the polyethylene block is divided into an upper and lower half, and each half is subdivided in six equal parts  along the longest symmetry axis. The twelve pieces are held together with eight stainless steel bolts.
\begin{table*}[h!]
\caption{Technical specifications of the NERO gas-filled proportional counters. (a) and (b) refer to the $^{3}$He detector models RS-P4-0810-104 and RS-P4-0814-207, respectively. \label{TabCounters}}
\begin{center}
\begin{tabular}{cccccc}
\hline \hline
 Detector       & Active   & Radius    & Nominal  & Gas                        & High      \\
                & Length   &           & Pressure & Composition                & Voltage   \\
                & (cm)     & (cm)      & (atm)    &                            & (+V)      \\ \hline
&  &  &  &  & \\
 $^{3}$He (a) & 25.0(2)    & 1.3(2)    & 10.2      & 100\% $^{3}$He             & 1350      \\
 $^{3}$He (b) & 35.6(2)    & 1.3(2)    & 4.0      & 100\% $^{3}$He             & 1100      \\
    BF$_{3}$  & 50.8(1)    & 2.5(2)    & 1.2      &$\textgreater$96\% $^{10}$B & 600       \\ \hline \hline
\end{tabular}
\label{tab:tubes}
\end{center}
\end{table*}



\subsection{Electronics}
\label{sec:electronics}
\begin{figure}[h!]
\begin{center}
\includegraphics[width=12cm]{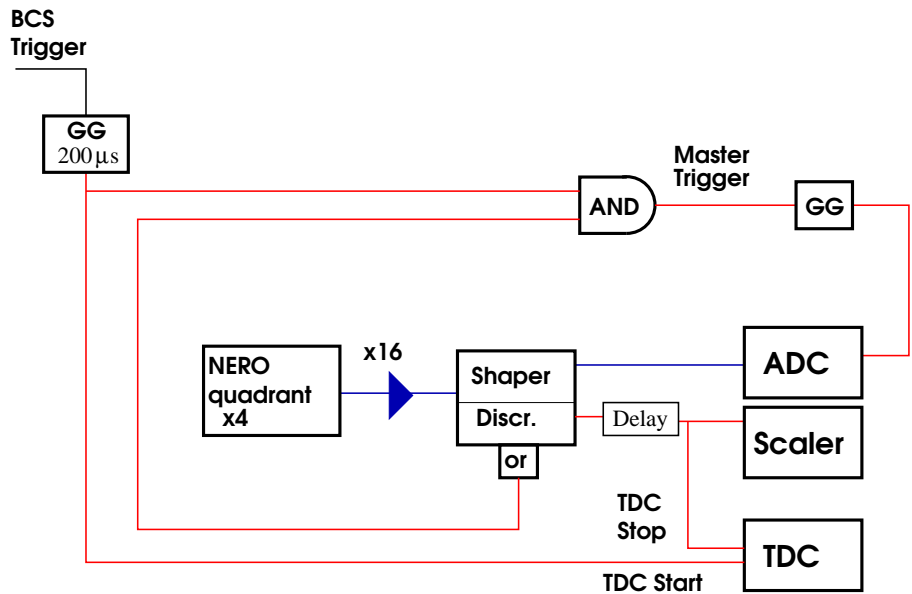}
\caption{NERO electronic diagram (see text for details). For clarity only one NERO quadrant is shown (GG stands for Gate Generator).}
\label{fig:electronics}
\end{center}
\end{figure}

The NERO readout channels are grouped in 4 quadrants with 15 channels each (4 $^3$He counters
and 11 BF$_3$ counters). Figure~\ref{fig:electronics} shows a schematic diagram of the NERO electronics for one quadrant. 
The proportional counters detect the charged particles produced in the exothermic neutron-capture reaction $^{3}$He($n$,$p$) or $^{10}$B($n$,$\alpha$), respectively.
Their signals feed 16-channel preamplifiers built at NSCL using Cremat CR-101D miniature charge-sensitive preamp chips. The pre-amplified signals are sent into four 16-channel shaper and discriminator modules, designed at Washington University, St. Louis, and manufactured by Pico Systems~\cite{Els09}. These modules integrate independent shaping and discriminating circuits sharing the same input. Shaping times and pole-zero cancelation are adjusted for each channel by properly selecting the capacitances. The gain and threshold levels of the shaper/discriminator are 
adjusted 
via computer control over the CAMAC bus.

The logic signals from the discriminator are recorded in scalers and
in a 64-channel multi-hit (VME) TDC that is common for all quadrants. The TDC was programmed to work in start-gate mode, in which a gate signal, generated by a $\beta$ decay detected in the BCS, enables the module to accept multiple stop signals in each channel from any of the sixty gas counters. The duration of this gate ($\tau$=200~$\mu$s) was chosen to account for the time needed to moderate and
detect the neutrons (see Sec.~\ref{sec:moderation}). The $P_{n}$ value of a given nucleus
is extracted from the number of stops-signals registered in the TDC
(i.e., neutrons correlated with $\beta$ decays) relative to the number of $\beta$ decays detected in the BCS.

\begin{figure}[t!]
\begin{center}
\includegraphics[width=4cm]{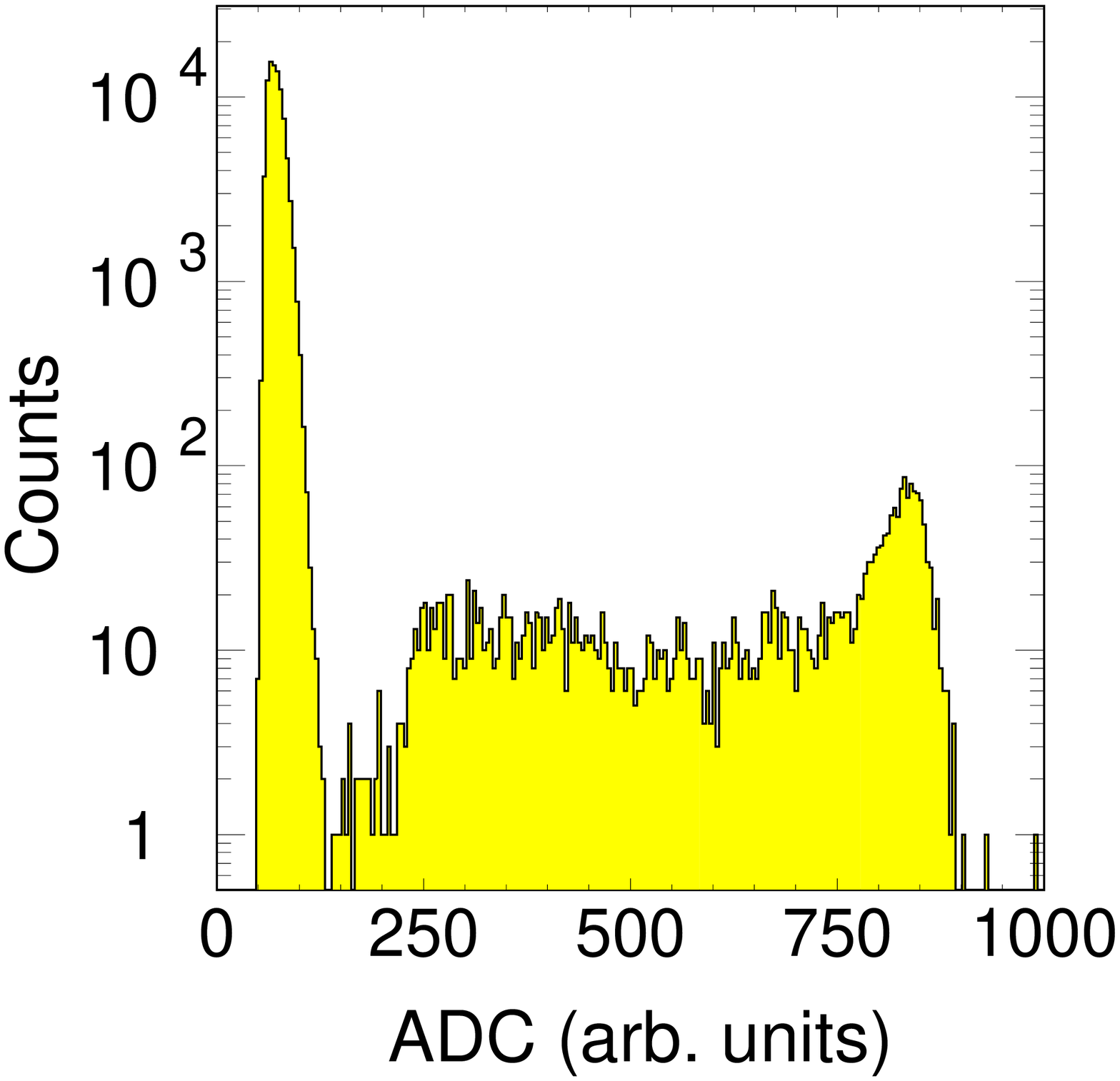}
\includegraphics[width=4cm]{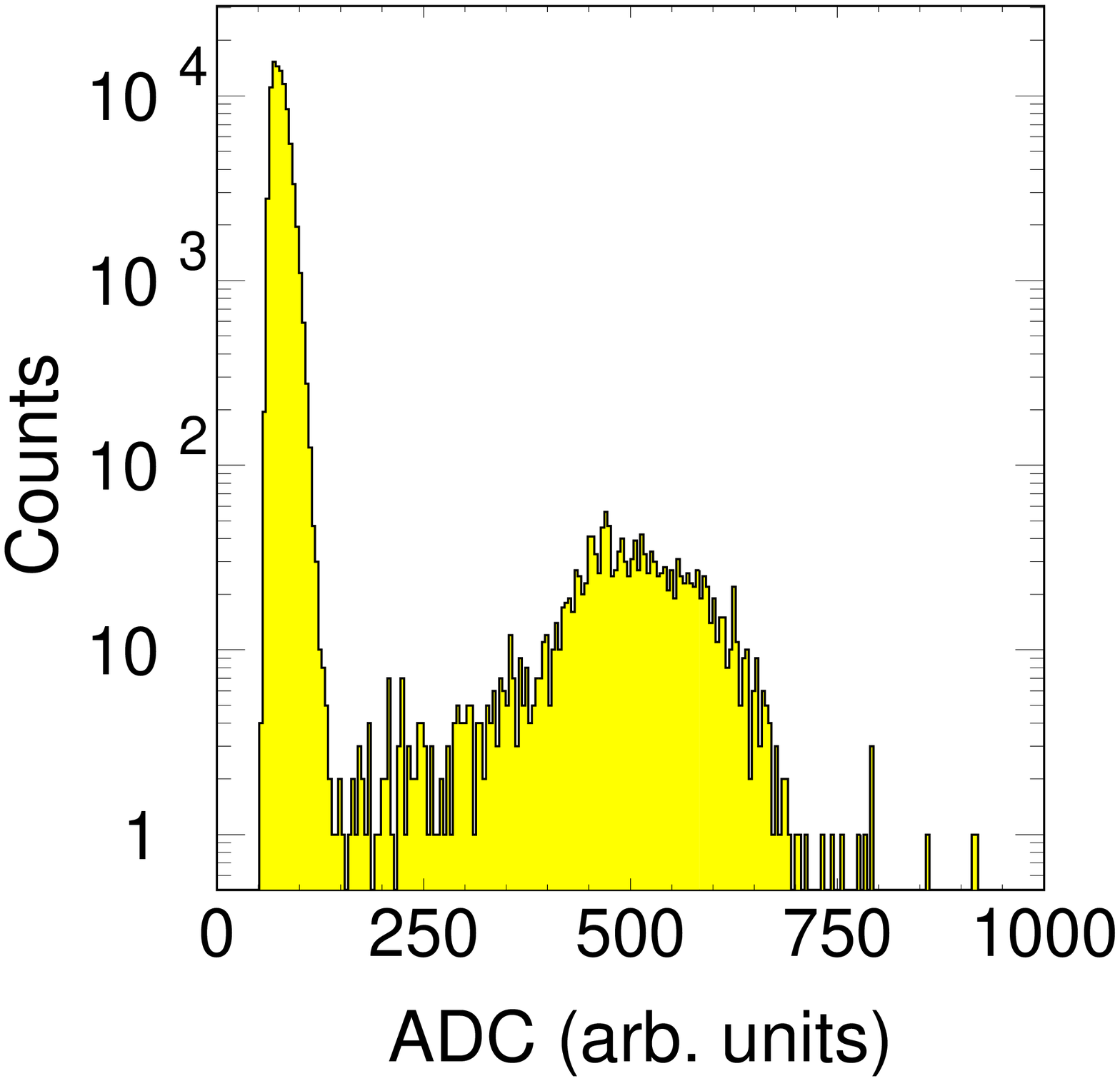} \\
\includegraphics[width=4cm]{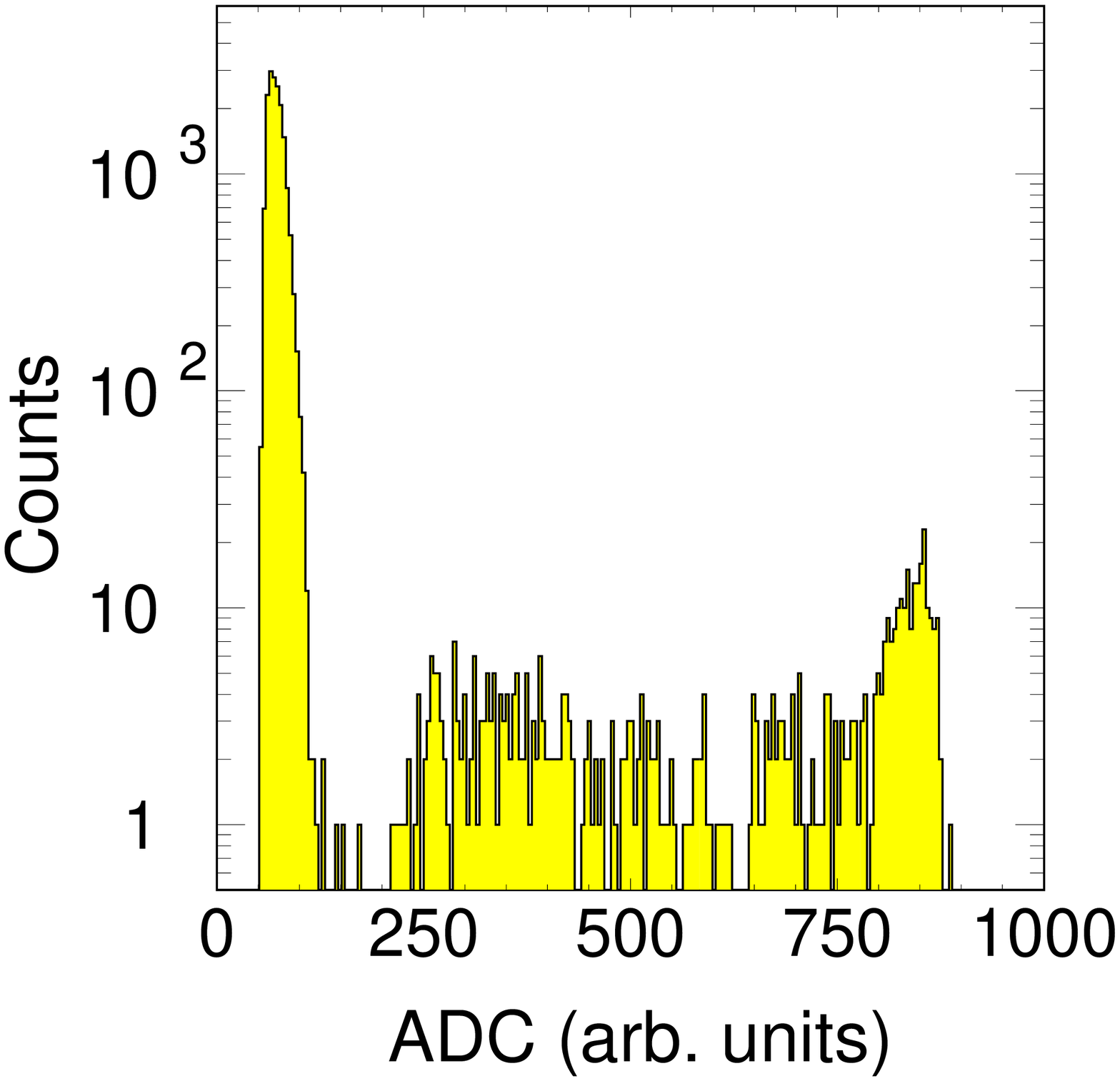}
\includegraphics[width=4cm]{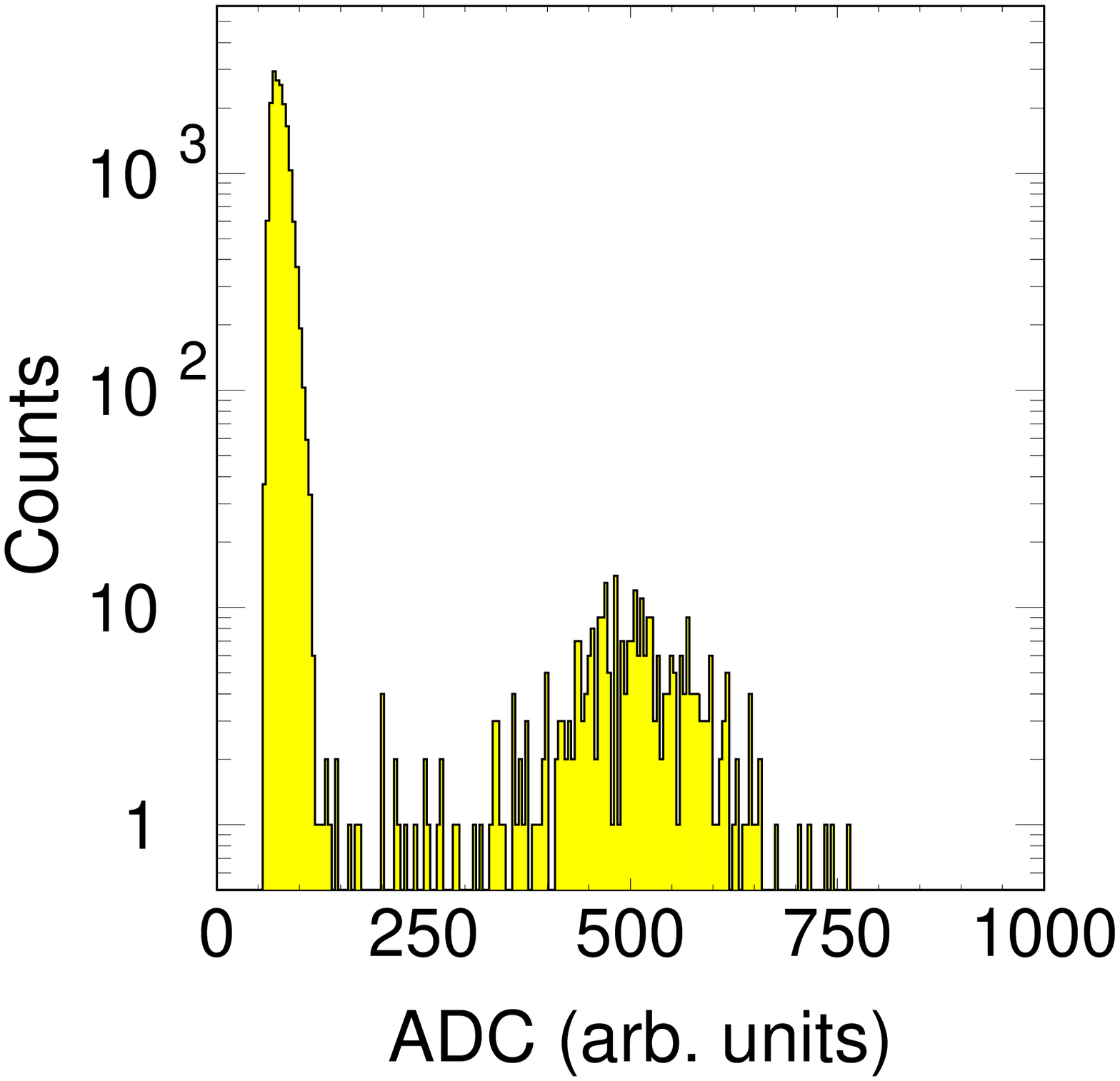} \\
\includegraphics[width=4cm]{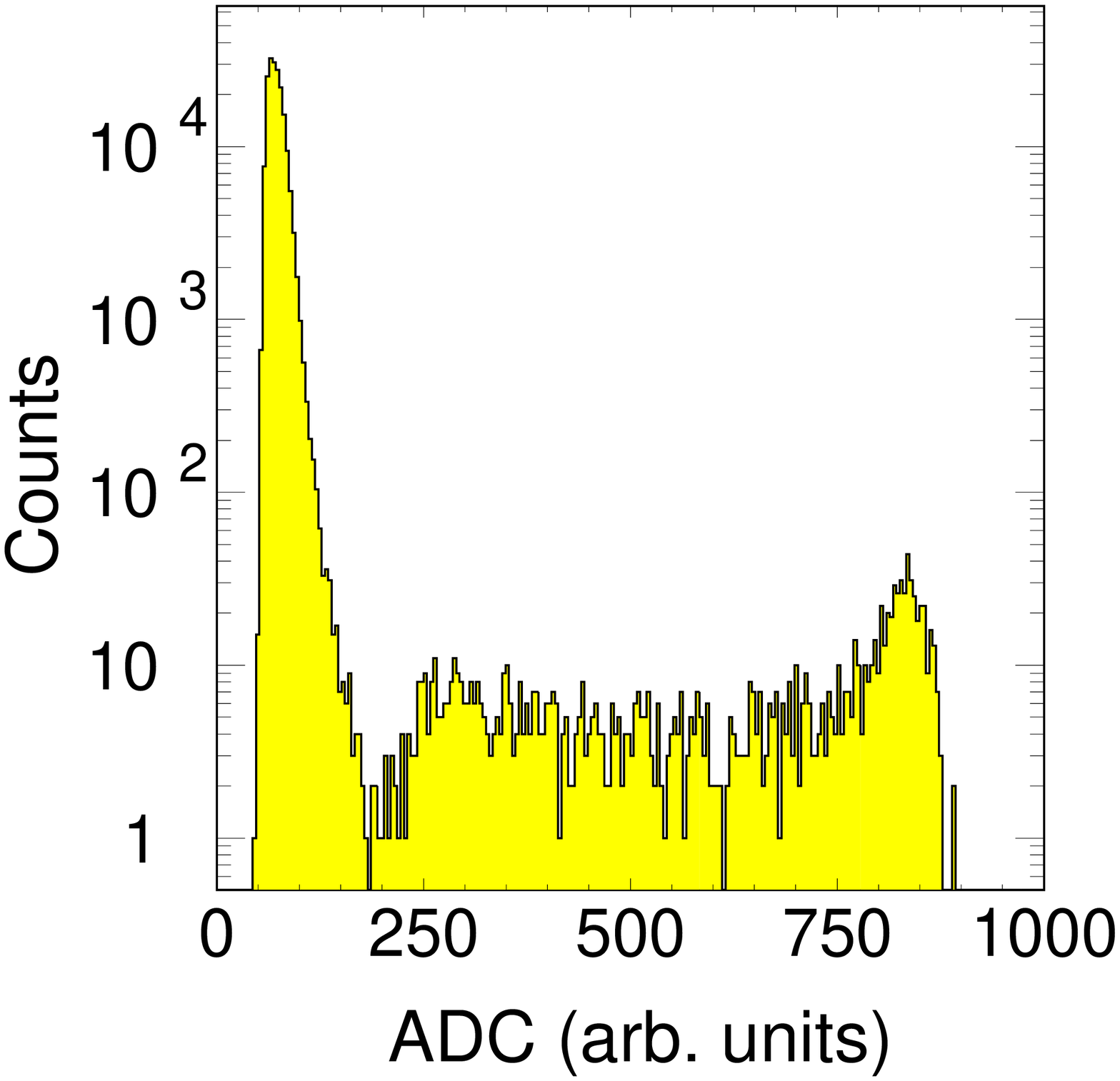}
\includegraphics[width=4cm]{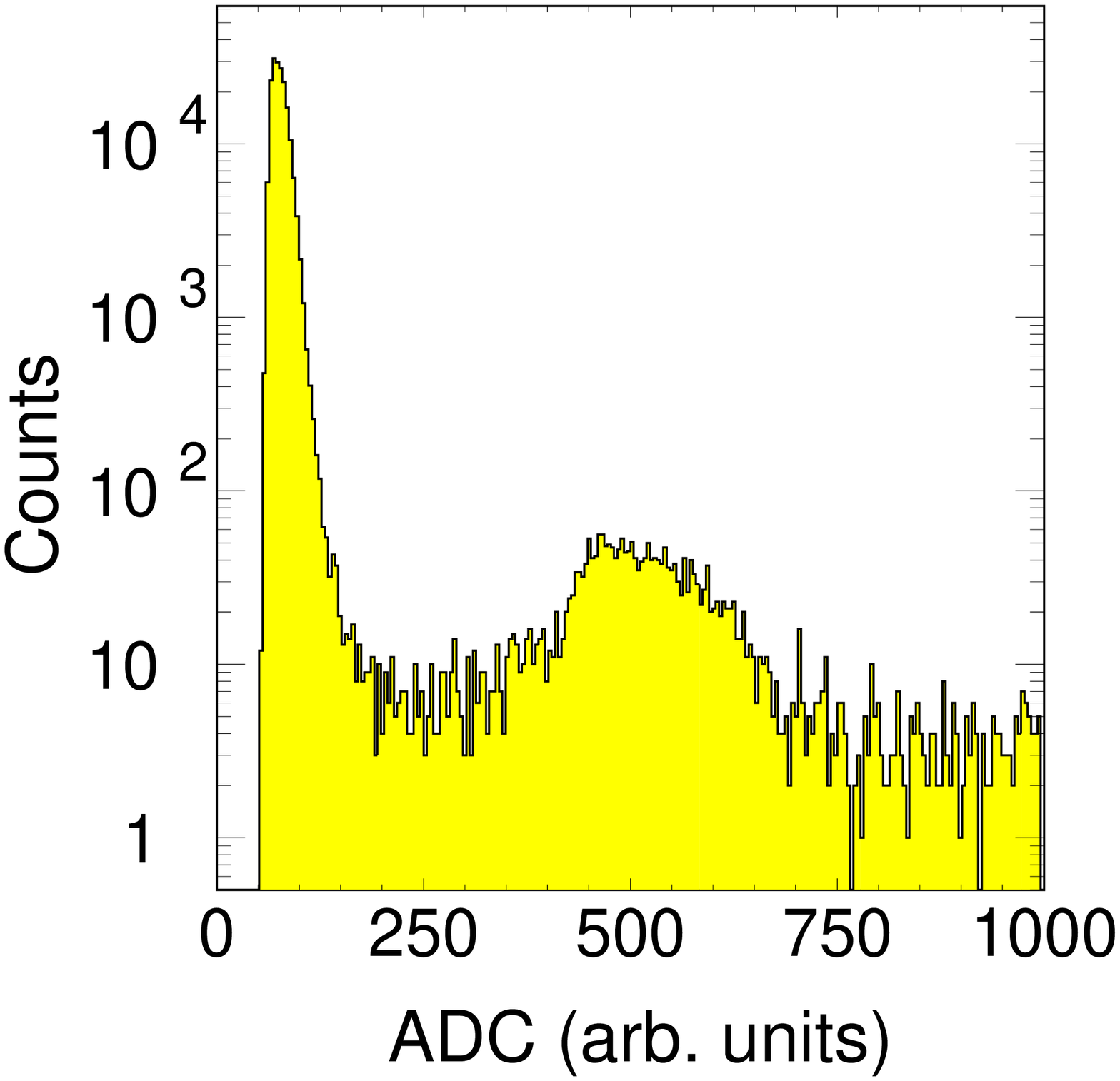} \\
\caption{NERO ADC spectra for one of the $^{3}$He (left panels) and BF$_{3}$ (right panel) gas counters. The top panels were recorded during 5 minutes, using a $^{252}$Cf neutron source at the center of NERO. The middle panels correspond to a one-hour measurement with nuclei produced in the reaction $^{136}$Xe (150~MeV/u)+Be that were implanted in the BCS and included
$\beta$ delayed neutron emitters. The data displayed in the bottom panels were recorded during a 12-hour background measurement without beam on target using NERO as trigger.}
\label{fig:ADC}
\end{center}
\end{figure}

The shaper outputs are connected to 32-channel (VME) ADC cards. 
The pulse height spectra recorded by the ADCs are used to set the thresholds and gains of the shaper/discriminator units, and to monitor any background or gain variation during the course of an experiment. Figure~\ref{fig:ADC} shows typical ADC spectra for $^{3}$He and BF$_{3}$ gas counters recorded under different conditions.  The spectra show the typical wall-effect: The location of the peak
at high amplitudes marks the Q value of the neutron-capture reaction, i.e., $^{3}$He($n$,$p$)$t$ and $^{10}$B($n$,$\alpha$)$^{7}$Li for the $^{3}$He and  BF$_{3}$ gas counters, respectively; the plateau or low energy tail at low amplitudes arises from events where reaction products hit the detector wall
preventing the complete deposition of their energy in the counter gas. Thresholds 
are set below these low amplitude events and just above the tail of the prominent low energy
peak generated by electronic noise and background $\gamma$ radiation.
Note that unlike the TDC,
the ADCs only register one neutron 
per BCS trigger.

\section{Detector performance}
\label{sec:performance}

\subsection{Moderation time}
\label{sec:moderation}
\begin{figure}[h!]
\begin{center}
\includegraphics[width=6.5cm]{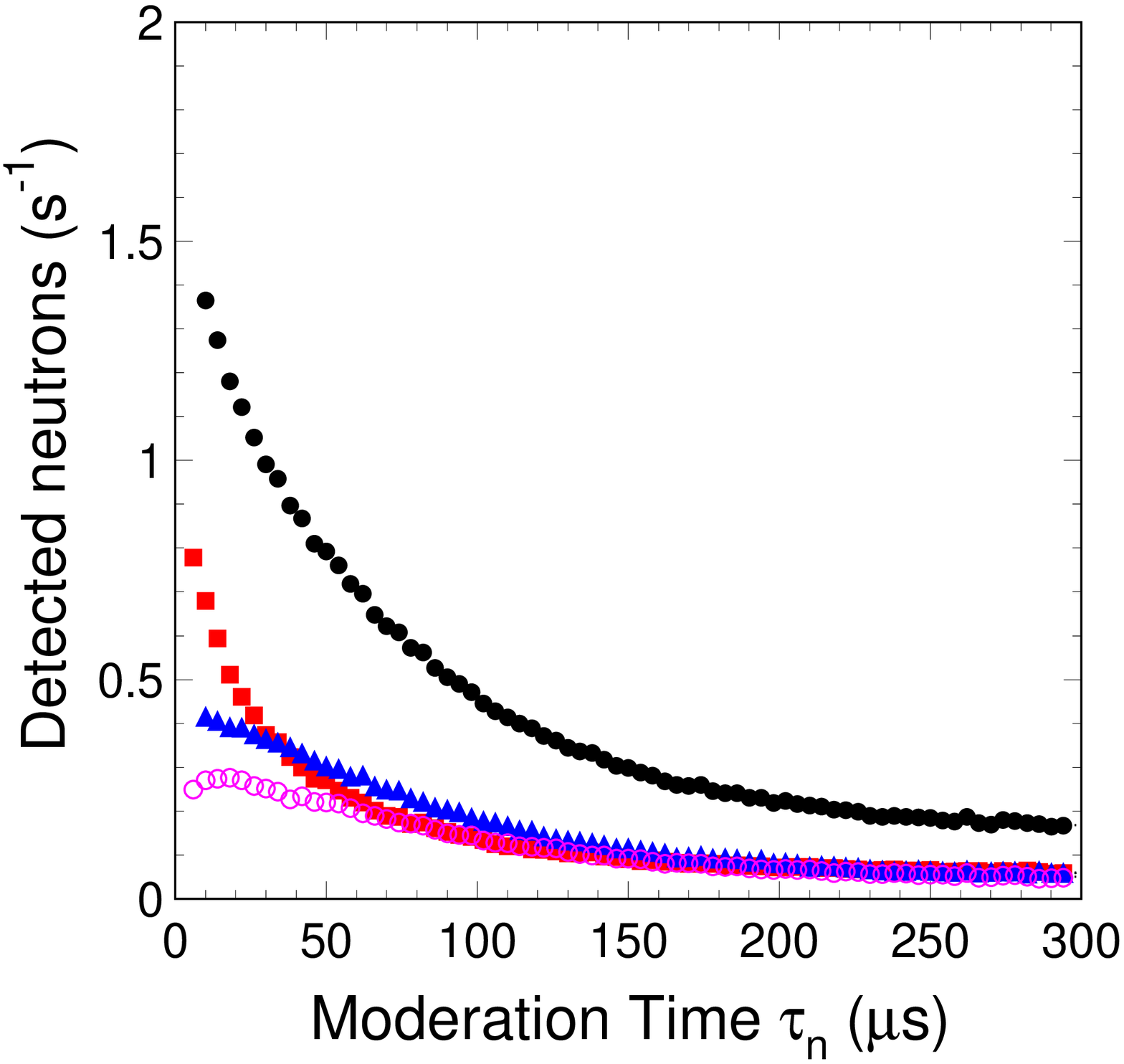}
\includegraphics[width=6.5cm]{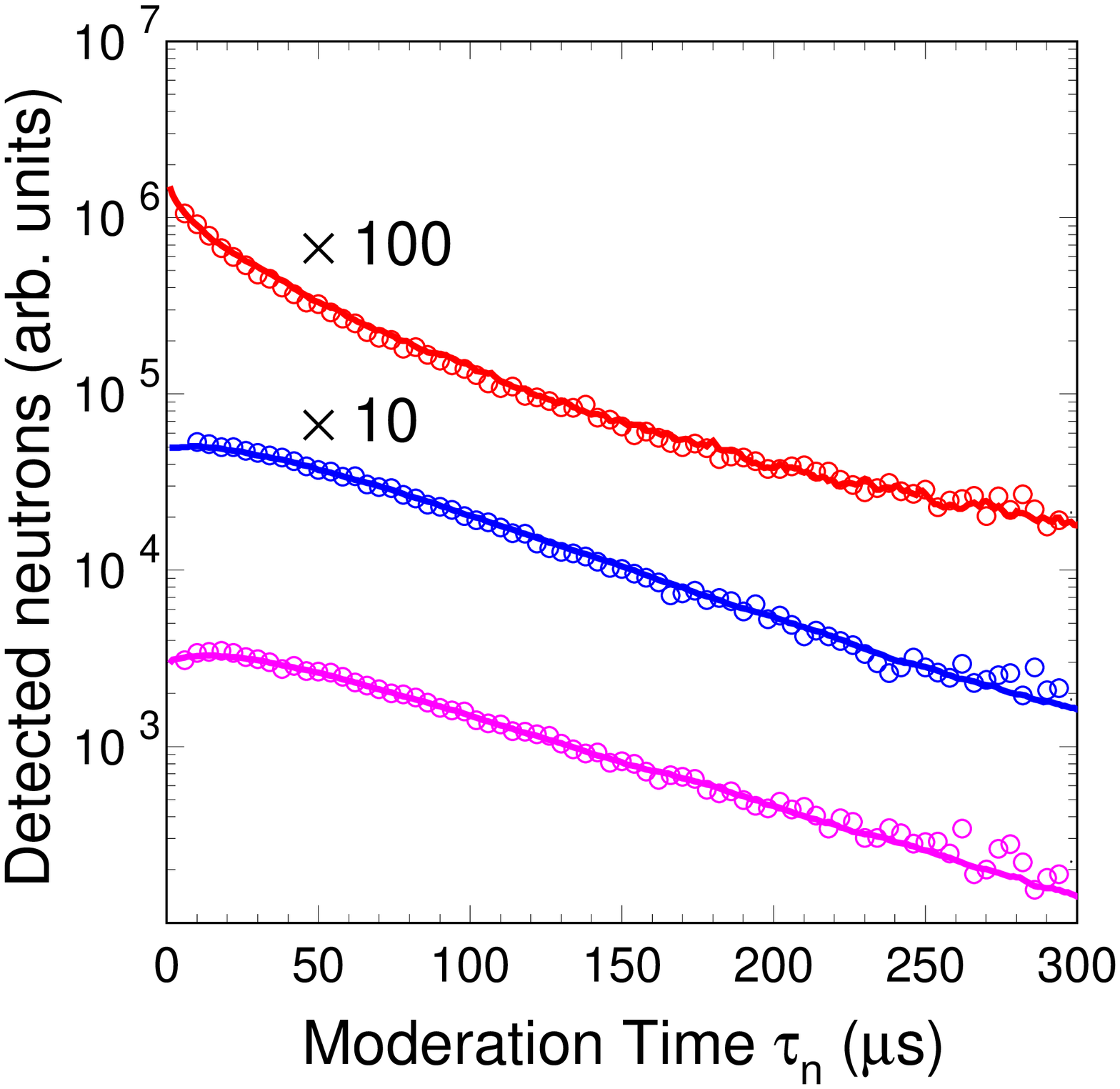}
\caption{(Color online). Left:
Measured background-subtracted moderation time distributions for a $^{252}$Cf source for the innermost ring (squares), the intermediate ring (triangles), the outer ring (empty circles) and for the entire detector (solid circles). Right: Moderation-time distribution for each ring (empty circles), compared with results obtained with MCNP (solid lines). For easy comparison, the measured and calculated distributions for the first and second rings are scaled by 100 and 10, respectively.
Note that the error bars of the experimental data are smaller than the symbol sizes.}
\label{fig:moderation}
\end{center}
\end{figure}

In order to determine the optimal duration of the TDC gate $\tau$, we measured the time needed by the neutrons to slow down in the polyethylene moderator before their detection ($\tau_{n}$). A $^{252}$Cf source was located at the center of NERO facing a NaI scintillator at a distance of 5~cm.
Neutrons and $\gamma$ rays were emitted in coincidence from the fragments produced in the spontaneous fission of $^{252}$Cf. The scintillator was used to detect the $\gamma$ rays, which provided the external trigger of the NERO electronics (replacing the BCS trigger shown in Fig.~\ref{fig:electronics}). The
time difference between the detection of a $\gamma$ ray and the subsequent moderated neutrons recorded in the multi-hit TDC provided $\tau_{n}$.
The distribution of $\tau_{n}$ is shown in Fig.~\ref{fig:moderation} (left) for the innermost, intermediate, and outer rings of proportional counters,  as well as for the whole detector.

The largest differences between the three rings are found at the shortest times, 
when most of the neutrons emitted from the center of NERO reach the innermost ring. At late times, the neutrons are more uniformly distributed over the whole moderator, and the three rings have similar detection rates.
For each ring, the excellent agreement of the measured, background-subtracted moderation-time distributions
with MCNP simulations is shown in Fig.~\ref{fig:moderation}.  Between 50~$\mu$s and 300~$\mu$s
the time distributions can be approximated with exponential functions. The corresponding measured
and calculated moderation time scales are  43~$\mu$s and 41~$\mu$s for the first ring, 51~$\mu$s and 52~$\mu$s for the second ring, and 55~$\mu$s and 59~$\mu$s for the third ring, respectively.

From Fig.~\ref{fig:moderation}, we find that 94.3(1)\% of the neutrons are detected within $\tau_{n}\leq$200~$\mu$s. The energies $E_{n}$ of the neutrons  emitted in the spontaneous fission of $^{252}$Cf are 
typically described by a Maxwell-Boltzmann distribution function with an effective temperature of $kT$=$1.42$~MeV~\cite{Bat05}, an average neutron energy of 2.1~MeV and a smooth tail at higher energies that extends up to 9~MeV. These are higher energies than typically expected for $\beta$-delayed neutrons (see discussion at the end of Sec.~\ref{sec:MCNPeff}).
Since the moderation time increases with neutron energy, we expect that
more than 94.3(1)\% of $\beta$-delayed neutrons are detected within $200$~$\mu$s.
We therefore chose a TDC gate of $\tau$=$200$~$\mu$s.

\subsection{Efficiency}
\label{sec:efficiency}

In order to characterize the NERO efficiency and its energy dependence, different types of measurements were performed using a $^{252}$Cf neutron source of known activity and neutrons produced in resonant and non-resonant reactions at the Institute for Structure and Nuclear Astrophysics (ISNAP) at the University of Notre Dame. The results obtained from these reactions were used to constrain the energy dependence of the NERO efficiency.

\subsubsection{Measurement of NERO efficiency with a $^{252}$Cf neutron source}
\label{sec:cfsource}
Before and after the calibration measurements performed at the University of Notre Dame,
the NERO efficiency was measured with a 1.251(5)~$\mu$Ci $^{252}$Cf
calibration source with an active diameter of 5~mm 
(neutron branching 11.6\% and half-life 2.689 years). Additional contributions to the total neutron rate from 
contaminants were estimated. Besides $^{252}$Cf, there are small impurities of $^{249-251}$Cf and $^{254}$Cf. $^{249}$Cf and $^{251}$Cf have a negligibly small spontaneous fission branch, whereas the present amount of $^{254}$Cf was very small due to its short half-life. Consequently, only the $^{250}$Cf and $^{252}$Cf isotopes had to be considered. At the time of the measurement, 3.6\% of the total activity of the source was due to $^{250}$Cf, whose contribution to the neutron activity was negligible due to its very low neutron branching of 
0.296\%. The alpha decay of $^{252}$Cf produces $^{248}$Cm, which  undergoes spontaneous fission accompanied with neutron emission, with a branching ratio of 8.39\%. The very long half-life of this radioisotope (3.48$\times$10$^{5}$ years) made its contribution to the total neutron rate negligible.

The number of detected neutrons was recorded with scalers and the multi-hit TDC described in Sec.~\ref{sec:electronics}. We verified that data processing dead time is negligible up to 50~KHz, well above
the activity of the source, using a random pulser. Similarly, the 2~$\mu$s dead-time in the proportional counters was negligible. Taking the ratio of the number of neutrons recorded with NERO to the number of neutrons emitted by the source, calculated from the known source activity and neutron branching,
we obtained a neutron detection efficiency of 
31.7(2)\%. This value also serves as a reference point to verify the NERO efficiency before, during, and after experiments.

\subsubsection{Measurement of NERO efficiency with resonant reactions}
\label{sec:resonant}
The two resonant reactions used to study the NERO efficiency were $^{13}$C($\alpha$,$n$)$^{16}$O~\cite{Bla73,Ram76,Bru93} and $^{11}$B($\alpha$,$n$)$^{14}$N~\cite{Lun80,Wan91}. The ISNAP KN Van de Graff
accelerated a beam of $\alpha$ particles impinging onto $^{13}$C and $^{11}$B targets of 14(2)~$\mu$g/cm$^{2}$ and 12$^{+4}_{-2}$~$\mu$g/cm$^{2}$ thickness, respectively, located at the center of the NERO symmetry axis. The rate of incident $\alpha$ particles ($I_{\alpha}$) was monitored with an isolated electron-suppressed plate behind the target. A total of two resonances for the reaction $^{13}$C($\alpha$,$n$)$^{16}$O, and one for the reaction $^{11}$B($\alpha$,$n$)$^{14}$N were used (see Table~\ref{tab:resonances}). Each resonance was completely mapped around its peak energy $E_{R}$ by detecting the number of neutrons as a function of $\alpha$-beam energy $E_{\alpha}$. A linear background function was fit underneath the resonance curve and used to subtract non-resonant
contributions and background neutrons. The efficiency of NERO was determined as the ratio of the number of detected neutrons to the number of neutrons $N_{n}$ produced in the resonant reaction.
Since the resonances considered here fulfill $\Gamma_{R} << \Delta E = E_{i} - E_{f}$ and  $\Gamma_{R} << E_{R}$ (see Table~\ref{tab:narrow}), one can use the thick-target narrow-resonance approximation and calculate $N_n$  (see for example \cite{Fow47}) using
\begin{equation}
N_{n} = \frac{I_{\alpha} t \pi^{2} \hbar^{2} (\omega \gamma)_{R} N_{A} \rho}{\mu A E_{R}}\left( \frac{dE}{dz} \right)^{-1}.
\label{eq:Nres}
\end{equation}
where $(dE/dz)$ is the stopping power of the $\alpha$ particle in the target material, calculated with the SRIM-2000 code~\cite{SRIM} in the center-of-mass frame; $t$ is the duration of the measurement; $A$ is the mole mass of the target; $\rho$ is the target mass density;
$\mu$ is the reduced mass of the system and $(\omega \gamma)_{R}$ is the resonance strength. 
The resonance parameters used in the calculations were taken from Refs.~\cite{Bla73,Ram76,Bru93,Lun80,Wan91}. They are summarized in Table~\ref{tab:resonances}.

The results of the efficiency measurements are shown in Table~\ref{tab:effreactions} for the three selected resonances, along with their corresponding evaluated average neutron energies in the 
laboratory frame $\langle E_{n} \rangle$. The calculation of this latter quantity is described in Sec.~\ref{sec:MCNPeff}.


\begin{table*}[h!]
\caption{Properties of the three selected resonances: resonance energy in the laboratory frame $E_{\alpha}$, resonance width $\Gamma_{R}$, strength $(\omega \gamma)_{R}$, excitation energy
 $E_{x}$, and spin and parity 
$J^{\pi}$.
}
\begin{center}
\begin{tabular}{cccccc}
\hline \hline
Reaction & $E_{\alpha}$ & $\Gamma_{R}$ & $(\omega \gamma)_{R}$   &  $E_x$  &  $J^{\pi}$   \\
         & (MeV)        & (keV)        & (eV)                    &  (keV)    &              \\ \hline
&  &  &  &  &  \\
 $^{13}$C($\alpha$,$n$)$^{16}$O & 1.053         & 1.5(2)                   & 11.9(6)         &  7165     & 5/2$^{-}$  \\
 $^{13}$C($\alpha$,$n$)$^{16}$O & 1.585         & $\leq$1                 & 10.8(5)         &  7576(2)  & 7/2$^{-}$  \\
 $^{11}$B($\alpha$,$n$)$^{14}$N & 0.606         & 2.5(5)$\times$10$^{-3}$ & 0.175(10)       & 11436     & 7/2$^{-}$  \\ \hline \hline
\end{tabular}
\label{tab:resonances}
\end{center}
\end{table*}

\begin{table*}[h!]
\caption{Validation of the thick-target, narrow-resonance approximation. $d \rho$ is the target
thickness and $\Delta E$ is the energy loss in the target at the resonance energy.
}
\begin{center}
\begin{tabular}{ccccc}
\hline \hline
 Reaction               & $E_{\alpha}$      &  $d \rho$             &   $\Delta E$  &  $\Gamma_{R}$                 \\
                        & (MeV)             &  (mg/cm$^{2}$) &   (keV)       &  (keV)                        \\ \hline
&  &  &  &  \\
 $^{13}$C($\alpha$,$n$)$^{16}$O & 1.053             &  0.014         &   18          &  1.5(2)                 \\
 $^{13}$C($\alpha$,$n$)$^{16}$O & 1.585             &  0.014         &   16          &  $\leq$1                      \\
 $^{11}$B($\alpha$,$n$)$^{14}$N & 0.606             &  0.012         &   18          &  2.5(5)$\times$10$^{-3}$      \\ \hline \hline
\end{tabular}
\label{tab:narrow}
\end{center}
\end{table*}

\subsubsection{Measurement of NERO efficiency with non-resonant reactions}
\label{sec:nonresonant}
Additional measurements of the NERO efficiency were performed at ISNAP using neutrons produced in the  $^{51}$V($p$,$n$)$^{51}$Cr reaction~\cite{Zys80} at three different energies. This reaction has been used in the past for neutron detector calibrations~\cite{Ram76,Lun80}. Here, a proton beam was accelerated at the KN accelerator and impinged onto a $^{51}$V target mounted in the center of NERO. Three incident proton energies of 1.8~MeV, 2.14~MeV and 2.27~MeV were chosen from regions of the excitation function with no individual resonances, using three targets with a thickness of
32~$\mu$g/cm$^2$.

To determine the number of  $^{51}$V($p$,$n$)$^{51}$Cr reactions that have occured
during a measurement one can take advantage of the fact that
for every $^{51}$V($p$,$n$)$^{51}$Cr reaction, a radioactive $^{51}$Cr 
is created with a half-life of 27.7025(24) days. The electron-capture decay of $^{51}$Cr 
is followed by the emission of several X-rays~\cite{Yal05} and a 320.0824(4)~keV $\gamma$ ray~\cite{Hel00} with a branching ratio of 9.91(1)\%~\cite{Yal05}. The number of neutrons $N_{n}$ produced in the reaction 
can then be simply determined from the activity of the target after irradiation.  The number of 320.1~keV $\gamma$ rays emitted was measured offline in a lead-shielded setup, where the irradiated target was mounted in a plastic holder facing a HPGe $\gamma$ detector. Decay losses during the irradiation, transport, and offline counting
were negligible.
The HPGe efficiency at 320.1~keV was found to be 0.76(4)\% using a $^{133}$Ba calibration source,
and by interpolating the efficiencies measured for the two $\gamma$ rays emitted at 302~keV and 356~keV.
The deduced NERO neutron efficiencies are listed in Table~\ref{tab:effreactions} for the three different proton energies. The systematic error is dominated by the 5\% uncertainty in the activity
of the $\gamma$-ray calibration source.

\begin{table*}[h!]
\caption{Laboratory frame projectile energy $E_{proj}$, average neutron energy $\langle E_{n} \rangle$, and width of the neutron energy distribution
 $\Delta E_{n}$ with corresponding measured
 neutron detection efficiency $\epsilon_n$.  The last three columns list efficiencies for isotropic sources emitting neutrons at the energy $\langle E_{n} \rangle$.
 $\epsilon_{n}$ \textquotedblleft isotr$\textquotedblright$
is derived from the experimental value $\epsilon_n$ using corrections calculated with MCNP,
 $\epsilon_{n}$ \textquotedblleft MCNP$\textquotedblright$ is the calculated efficiency, and
$\epsilon_{n}$ \textquotedblleft MCNP scaled $\textquotedblright$ is the calculated efficiency scaled to
obtain a best fit to the experimental data.}
\begin{center}
\begin{tabular}{cccccccc}
\hline \hline
Reaction               & $E_{proj}$ &  $\langle E_{n} \rangle$  & $\Delta E_{n}$ & $\epsilon_{n}$ & $\epsilon_{n}$  & $\epsilon_{n}$ & $\epsilon_{n}$ \\
                       &            &                           &      &        &      isotr.     &    MCNP        & MCNP           \\
                       &            &                           &      &                &                 &                & scaled         \\
                       & (MeV)      &    (MeV)                  & (MeV)&    (\%)        &       (\%)      &     (\%)       & (\%)           \\ \hline
                       &            &                           &      &                &                 &                &                \\
$^{11}$B($\alpha$,$n$) &  0.606     &                   0.56     & 0.12  & 33(2)          & 38(2)           & 42             &  37            \\
$^{13}$C($\alpha$,$n$) &  1.053     &                   2.8     & 0.31  & 24(1)          & 30(1)           & 29             &  26            \\
$^{13}$C($\alpha$,$n$) &  1.585     &                   3.2     & 0.41  & 27(1)          & 33(1)           & 29             &  24            \\
$^{51}$V($p$,$n$)      &  1.80      &                   0.23    & 0.014 & 39(2)          & 36(2)           & 44             &  39            \\
$^{51}$V($p$,$n$)      &  2.14      &                   0.55    & 0.024 & 34(2)          & 32(2)           & 42             &  37            \\
$^{51}$V($p$,$n$)      &  2.27      &                   0.68    & 0.028 & 34(2)          & 34(2)           & 41             &  36            \\ \hline \hline
\end{tabular}
\label{tab:effreactions}
\end{center}
\end{table*}

\subsubsection{Results and discussion}
\label{sec:MCNPeff}
\begin{figure}[h!]
\begin{center}
\includegraphics[width=6.5cm]{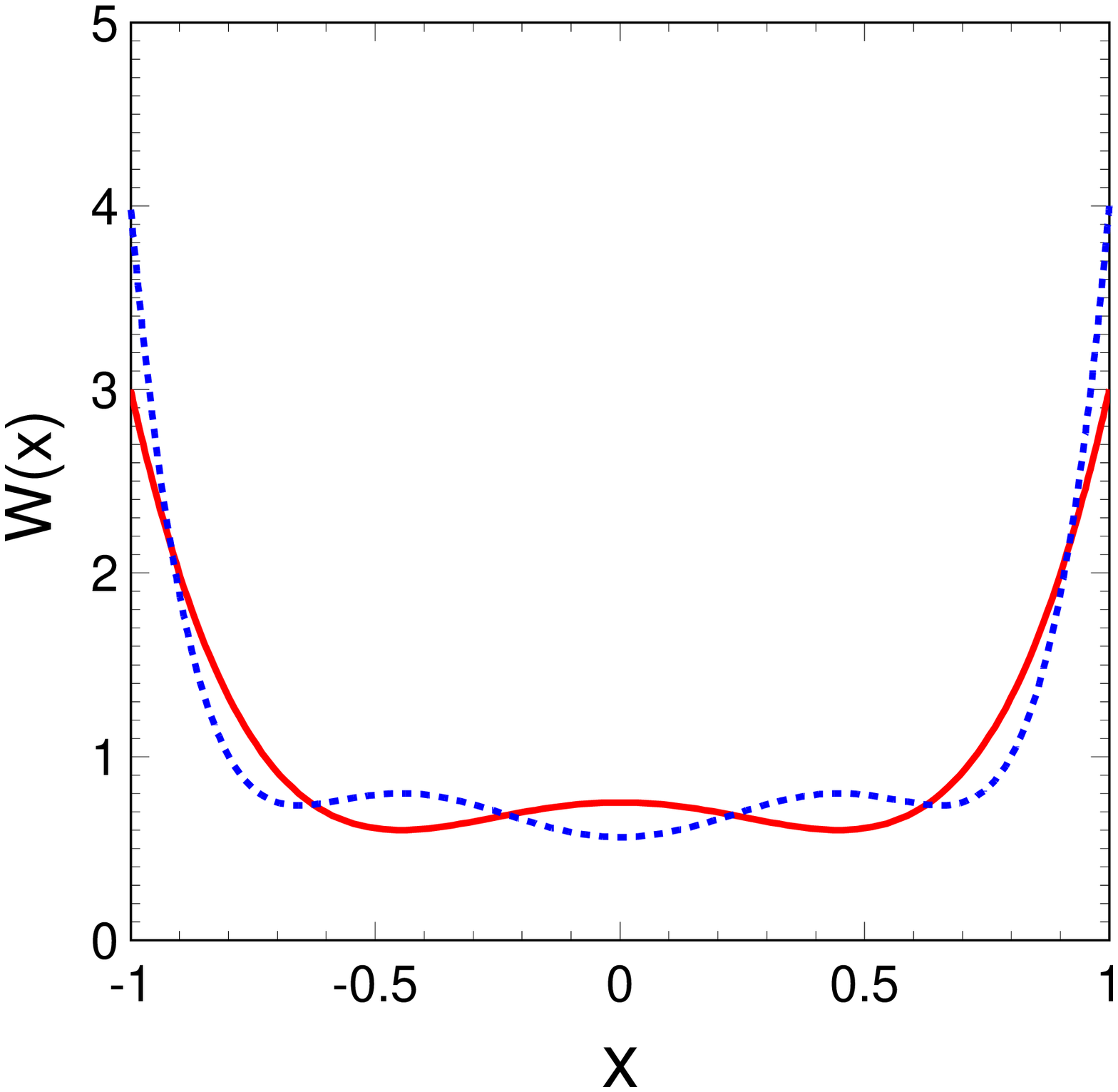}
\includegraphics[width=6.5cm]{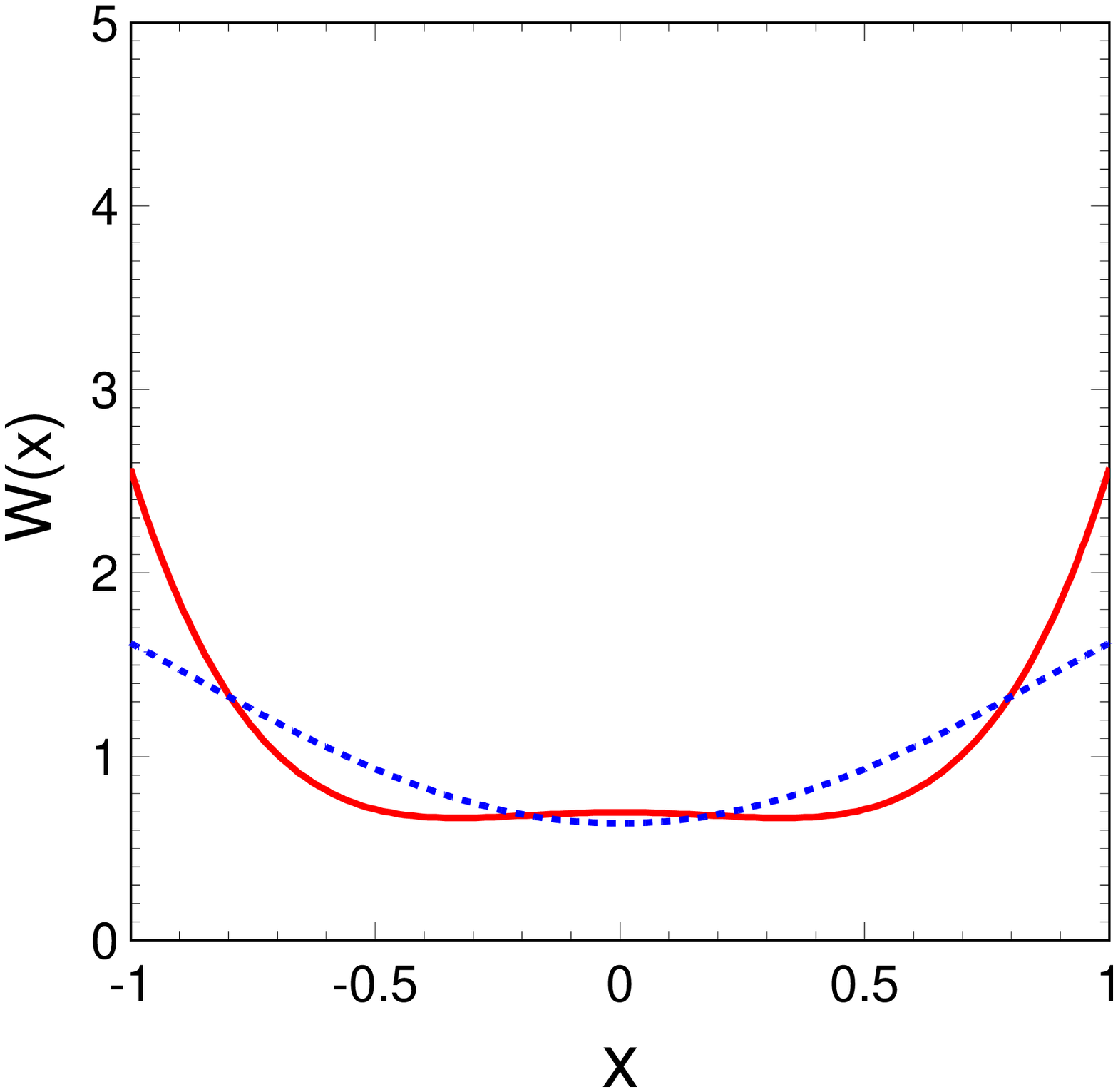}
\caption{Left: Angular correlation $W(x)$ calculated in the center-of-mass frame for the 7.165~MeV, $5/2^{-}$ (solid line) and 7.576~MeV, $7/2^{-}$ (dashed line) resonances in the reaction $^{13}$C($\alpha$,$n$)$^{16}$O. Right: Angular correlation $W(x)$ calculated in the center-of-mass frame for $J_{f}$=$1/2^{+}$ (solid line) and $J_{f}$=$3/2^{+}$ (dashed line) of the 11.436~MeV, $7/2^{-}$ resonance in the resonant reaction $^{11}$B($\alpha$,$n$)$^{14}$N. 
}
\label{fig:W}
\end{center}
\end{figure}

In order to 
evaluate the energy-dependence of the efficiency, 
the energy spectrum of the emitted neutrons
needs to be known for each reaction used. 
These neutrons are, for our purposes, essentially mono-energetic in the center-of-mass frame. The center-of-mass energies $\hat{E}_{n}$ can be calculated 
from the known reaction Q-values.
However, the corresponding laboratory frame neutron energy $E_{n}$ depends on the 
center-of-mass polar angle $\hat{\theta}$ of the emitted neutron with respect to the beam axis. 
This leads to a broadening of the neutron
energy distribution, with an average neutron energy 
$\langle E_{n} \rangle$ of
\begin{equation}
\langle E_{n} \rangle = \int_{-1}^{+1} E_{n}(x) W(x) dx,
\label{eq:avEn}
\end{equation}
where $x=\cos{\hat{\theta}}$. The angular-correlation functions $W(x)$ for each of  the three
resonances used here were calculated as described in  \cite{Iliadis08} and are shown in
Fig.~\ref{fig:W}. In the case of the $^{11}$B($\alpha$, $n$)$^{14}$N reaction, the coupling of the neutron spin with the ground-state of $^{14}$N leads to two possible values of the final spin. Variations of the results obtained using the two possible $W(x)$ functions 
were included in the final uncertainty. 
For $^{51}$V($p$,$n$)$^{51}$Cr we have chosen energies where the excitation function shows
non resonant behavior. This
justifies the assumption of isotropic neutron emission in the center-of-mass frame, or, equivalently, $W(x)=1/2$.

\begin{figure}[t!]
\begin{center}
\includegraphics[width=6.5cm]{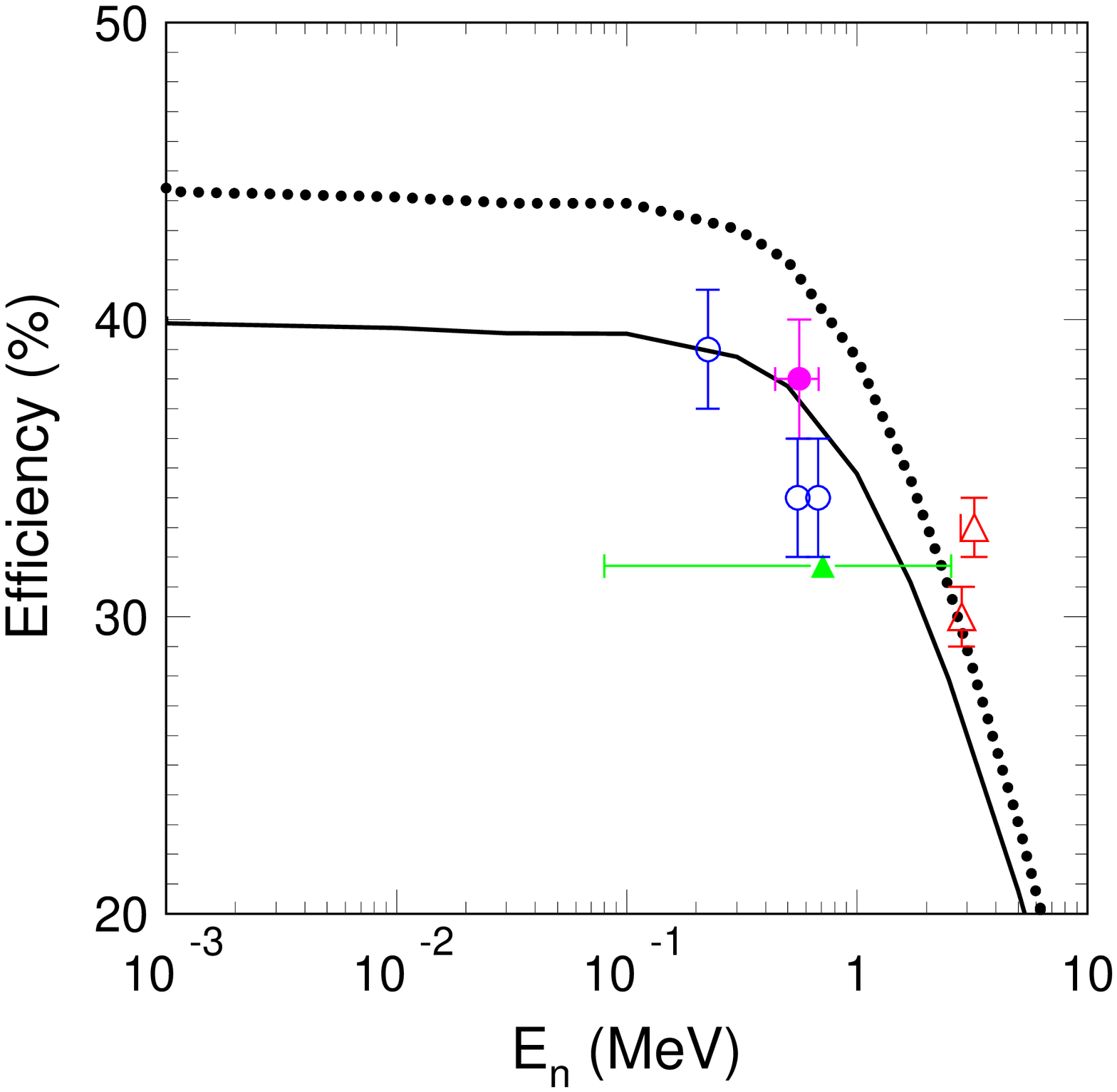}
\includegraphics[width=6.5cm]{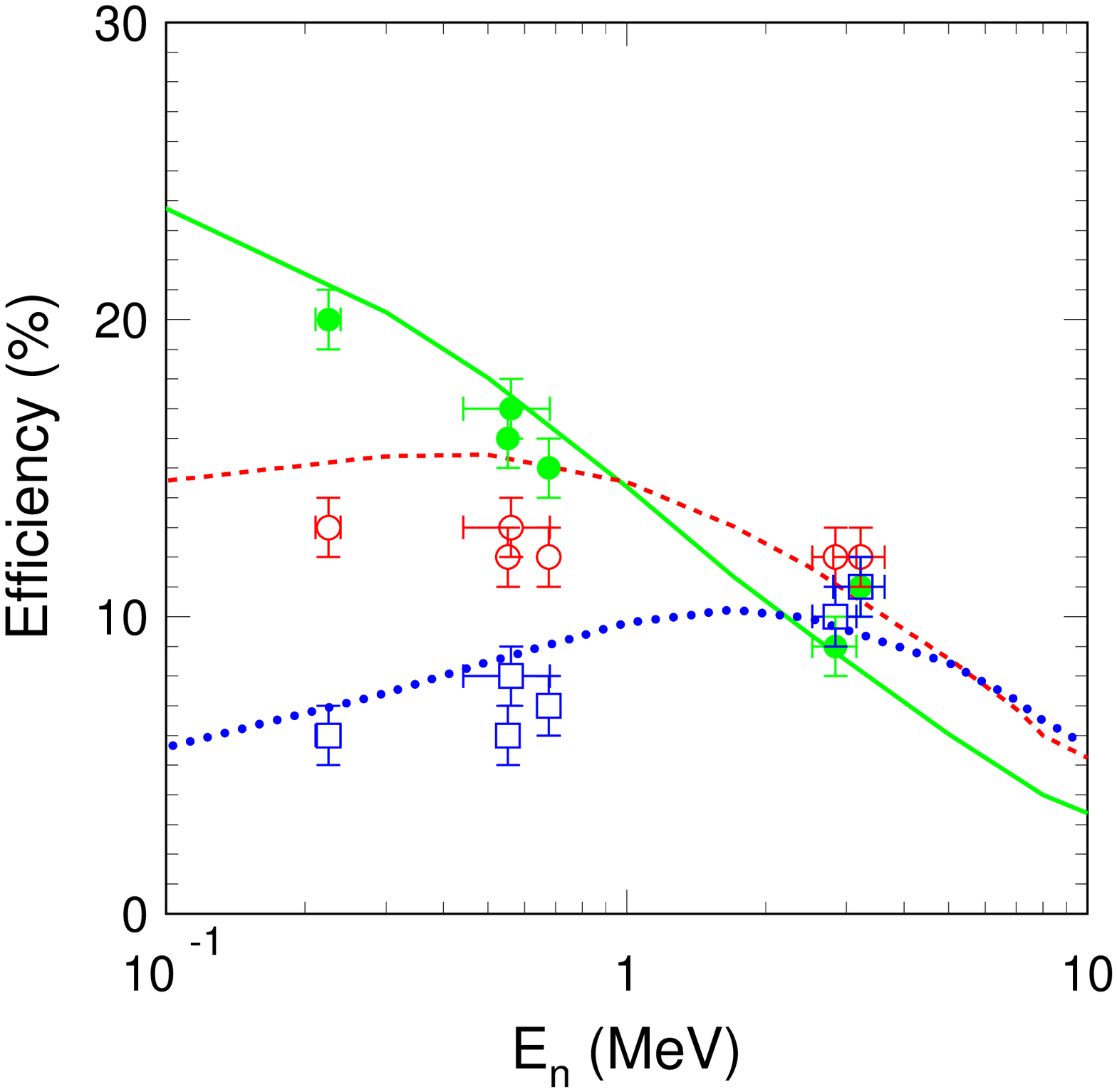}
\caption{(Color online). Left: MCNP-calculated total efficiency as a function of the neutron energy $E_{n}$ scaled to the experimental data (solid line) and unscaled (dotted line), compared with measurements for the reactions $^{11}$B($\alpha$,n) (solid circle), $^{13}$C($\alpha$,n) (empty triangles), $^{51}$V(p,n) (empty circles), and with the $^{252}$Cf neutron source (solid square). The energy width of the $^{252}$Cf measurement was calculated according to the shortest-interval criterion defined by W.~Br\"uchle for asymmetric distributions~\cite{Bru03}. 
Right: MCNP-calculated efficiencies as a function of neutron energy $E_{n}$ for the innermost ring (solid line), intermediate ring (dashed line) and outer ring (dotted line), compared to the measured values for the first (solid circles), second (empty circles) and third (empty squares) rings. Note the different 
scales of the two figures.}
\label{fig:NEROeffRing}
\end{center}
\end{figure}

The average neutron energies calculated in the laboratory frame $\langle E_{n} \rangle$ together with
the  width of the energy distribution $\Delta E_{n}$ are shown in Table~\ref{tab:effreactions} for each reaction, along with the corresponding measured efficiencies $\epsilon_{n}$. In the case of
$^{51}$V($p$,$n$)$^{51}$Cr $\Delta E_{n}$ also includes a small contribution from the energy loss and
straggling in the target.
In order to determine the efficiencies for neutrons emitted isotropically with a given energy  from the DSSD catcher in the BCS, our measured $\epsilon_{n}$ need to be corrected for the angular distribution and energy range of the neutrons 
(see e.g. Ref.~\cite{Sch57} for the reaction $^{13}$C($\alpha$,$n$)$^{16}$O).
This was done 
using MCNP simulations: First, the NERO efficiencies were calculated with 
MCNP 
at the energies $\langle E_{n} \rangle$ of Table~\ref{tab:effreactions}, assuming an isotropic mono-energetic neutron source located in the center of the detector. A second calculation
was then performed,  using the calculated laboratory frame angular and energy distributions of the neutrons.
The MCNP-calculated 
anisotropic-to-isotropic efficiency ratios
were then used as a correction factor to
translate the measured efficiencies into efficiencies for isotropic emission at a single energy
 $\langle E_{n} \rangle$. In Table~\ref{tab:effreactions}, we show the measured efficiencies for the different reactions ($\epsilon_{n}$) and the corresponding corrected values ($\epsilon_{n}$ isotr.). The strongest correction of about 15$\%$ arises mainly from the angular correlation $W(x)$ in the resonant reactions. The  isotropic efficiencies are shown in Fig.~\ref{fig:NEROeffRing} for the whole detector (left), and for each ring separately (right).

The experimentally-determined efficiencies covered a range of energies from about 0.2~MeV to 3~MeV. In order to extrapolate the results to lower energies, we combined the experimental values ($\epsilon_{n}$ isotr.) with a MCNP calculation of the efficiency as a function of energy
(see dotted line in Fig.~\ref{fig:NEROeffRing}, left). Despite a global absolute overestimation of about 5\%, the calculated $\epsilon_{n}(E_{n})$ function follows very well the energy dependence obtained from the measured data. 

As shown in Fig.~\ref{fig:NEROeffRing} (right), 
the efficiencies calculated independently for each ring follow reasonably well the 
energy trends of the experimental data.
For all the reactions investigated, MCNP reproduces the efficiency of the innermost ring, which is the most efficient of the three. The agreement is somewhat worse for the other rings at the lowest energies. In particular, the calculations overestimate the efficiency of the middle ring by about 3\% at energies below 700~keV.
We investigated the possibility that this discrepancy
could be related to the 
type of detectors used. 
Several 
test measurements of the efficiency were performed with a $^{252}$Cf source, using one single $^{3}$He proportional counter placed in the first, second and third ring, using the $^{252}$Cf source. A comparison of the calculated efficiencies with the values measured under these conditions was consistent with the results shown
in Fig.~\ref{fig:NEROeffRing} (right).
The overall 5$\%$ absolute overestimation of the efficiency by the calculations
could not be attributed to the uncertainty in the polyethylene density.
First of all, variations of the density modified the calculated efficiencies for the second and third ring
in the opposite direction of the first ring. Secondly, when the variations in density were limited to the uncertainties provided by the supplier, no differences in the calculated results were observed.

The good agreement of MCNP with experimental data observed in Fig.~\ref{fig:NEROeffRing} (right) for the first NERO ring with a thinner moderator layer, 
and the small discrepancies found  for 
the second and third rings with thicker moderator layers points to a limitation of MCNP to accurately calculate the scattering process of the neutrons in the moderator material. One
possibility would be molecular vibrational and rotational excitation modes in the moderator material. Whereas this problem would be hardly observable in detectors with thin moderators (e.g. Ref.~\cite{Dan95}), it would become 
more severe for thicker moderators. 
Interestingly, similar conclusions were drawn when comparing MCNP calculations with neutron-flux measurements performed with thick neutron detectors~\cite{ESARDA,Wol99}.
In order to compensate for these model deficiencies we scaled the
calculated efficiencies for each ring independently to better match the
experimental data (see scaled efficiency in Table~\ref{tab:effreactions}). The new scaled efficiency-curve (solid line in Fig.~\ref{fig:NEROeffRing}, left) can thus be used to extrapolate the efficiency to energies below 200~keV.

It is worth noting that the relevant neutron energy range for $\beta$-delayed neutron emission
in r-process nuclei 
is a few hundred keV.
As an example, for the r-process nuclei around $A \sim 100-130$, spectroscopic studies of $\beta$-delayed neutron emitters~\cite{Kra79a, Kra79b, Kra82} showed that $\hat{E}_{n}$ is typically much lower than $Q_{\beta}-S_{n}$. Neutron energies in the laboratory frame $E_{n}$ were found to
be 199~keV for $^{87}$Br, 450~keV for $^{98}$Rb, and 579~keV for $^{137}$I. This result was further supported by the measured average neutron energies of fission fragments from $^{235}$U ($\langle E_{n} \rangle=$575~keV) and $^{239}$Pu ($ \langle E_{n} \rangle=$525~keV), where, in addition, very few neutrons were found at $E_{n}$$\gtrsim$800~keV~\cite{Kra79a,Eng88}. According to these authors, the reason for the \textquotedblleft compressed$\textquotedblright$ $E_{n}$ spectra is the preferred population of the lowest excited states in the final nuclei~\cite{Kra82}. Our experimental efficiency calibration therefore covers the most critical energy range, and the condition of an energy independent
efficiency for $\beta$-delayed neutrons is well fulfilled.
As an example, $\epsilon_{n}$ shown in Fig.~\ref{fig:NEROeffRing} (left) shows a relative
variation of about $\pm$5\% for energies below 800~keV. This variation will contribute to the
final uncertainty of the measured $P_{n}$.
This uncertainty can be reduced if the neutron energies $E_{n}$ can be constrained from experiment or theory.

\subsection{Background}
\label{sec:background}
One limitation for the measurement of $P_{n}$, particularly for very exotic nuclei, is the neutron background rate ($B_{n}$). Its estimation requires to distinguish two different origins. First, there is the \textquotedblleft intrinsic$\textquotedblright$ background 
associated with the electronics of the NERO detector and its sensitivity to the neutrons present in the environment (mainly cosmic rays). Secondly, during the course of an experiment, there are beam-induced neutrons produced by nuclear reactions unrelated to the $\beta$-delayed neutron
emission of interest. During experiments, neutron background rates measured with NERO in
self-trigger mode can vary within about 5--10~$s^{-1}$ 
depending on whether or not the beam is on target. As will be discussed later, the impact
of these background rates is dramatically reduced when the neutrons are measured in coincidence with $\beta$ decays.

\begin{figure}[t!]
\begin{center}
\includegraphics[width=4.2cm]{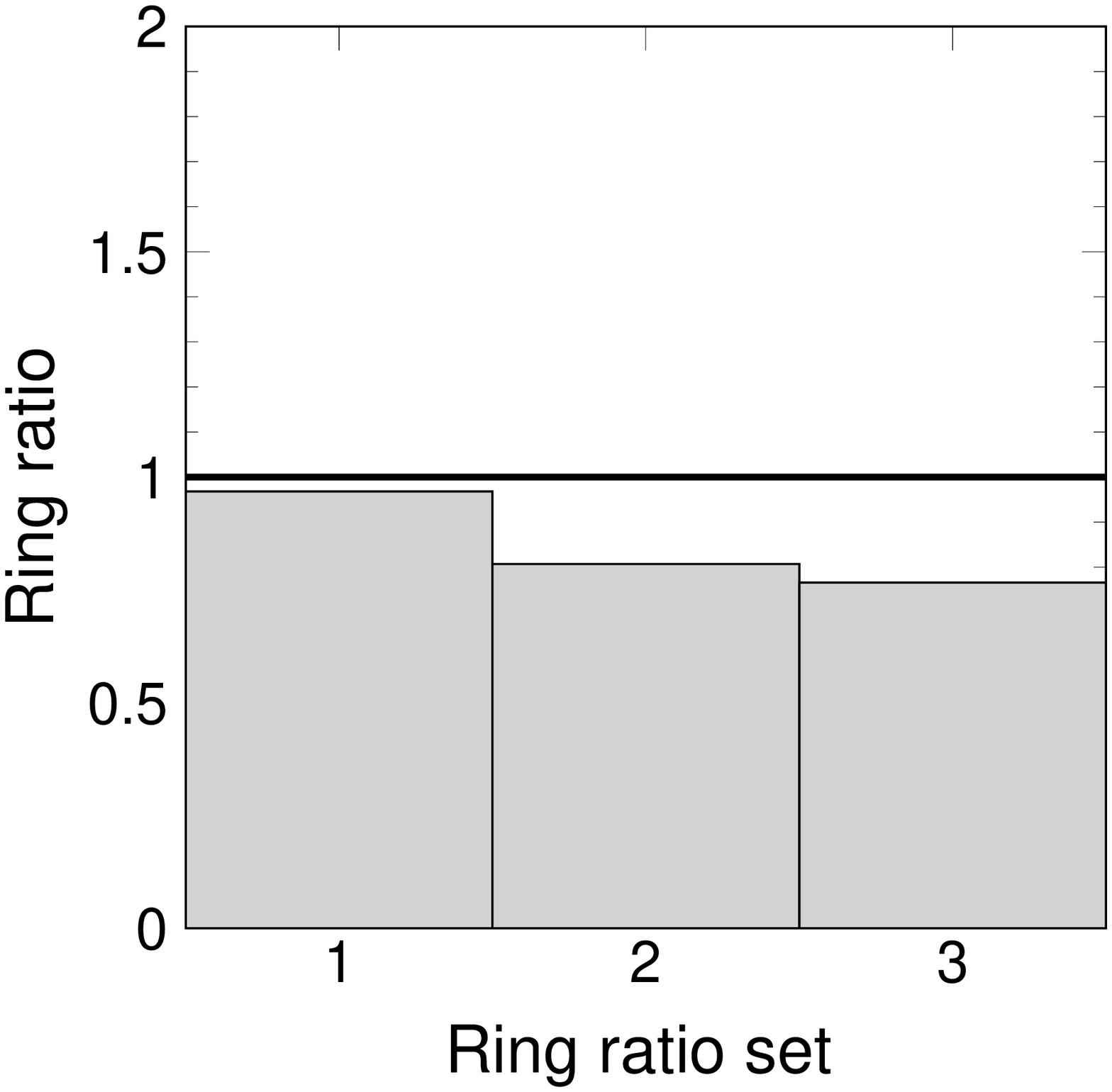}
\includegraphics[width=4.2cm]{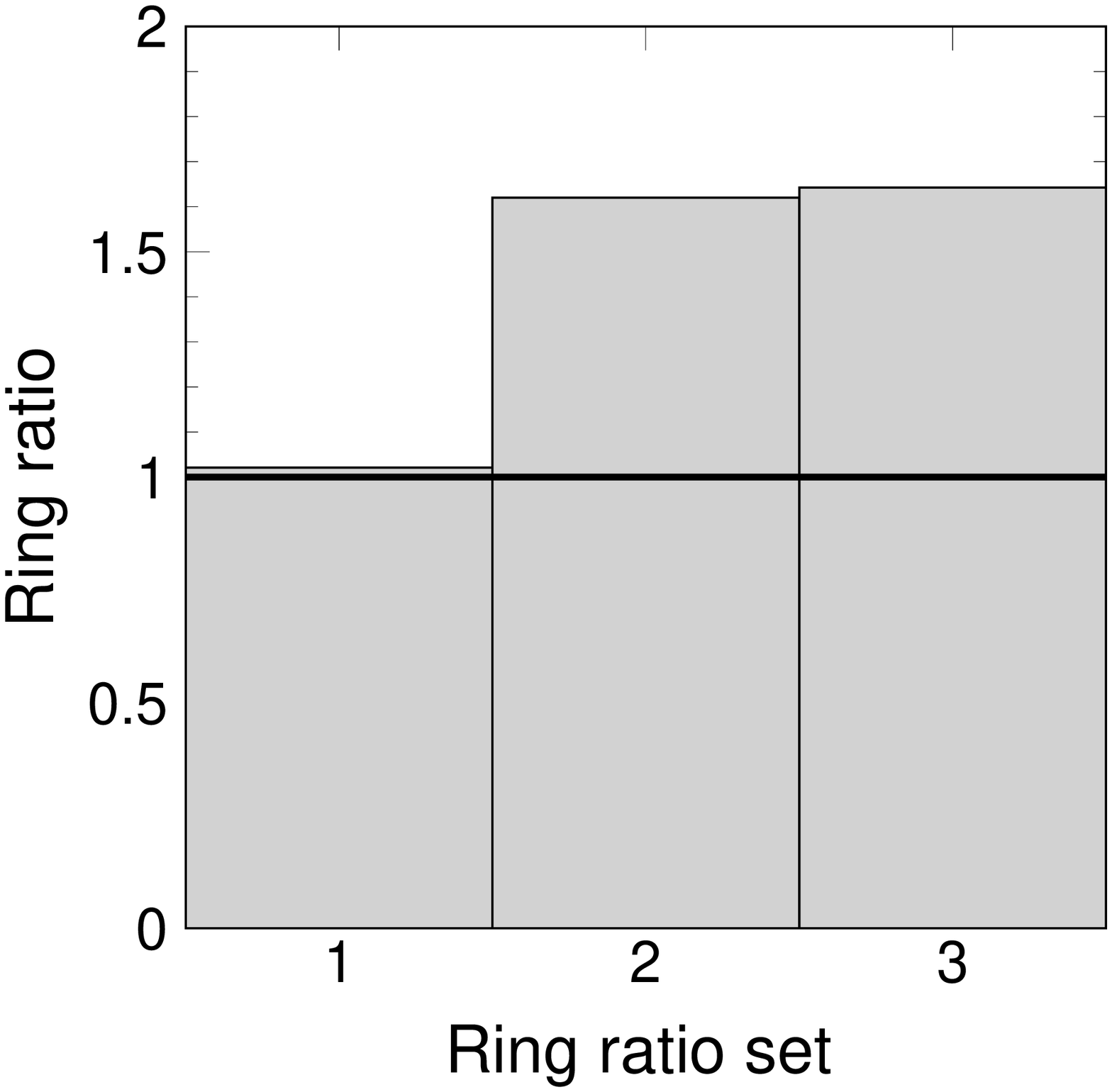}
\includegraphics[width=4.2cm]{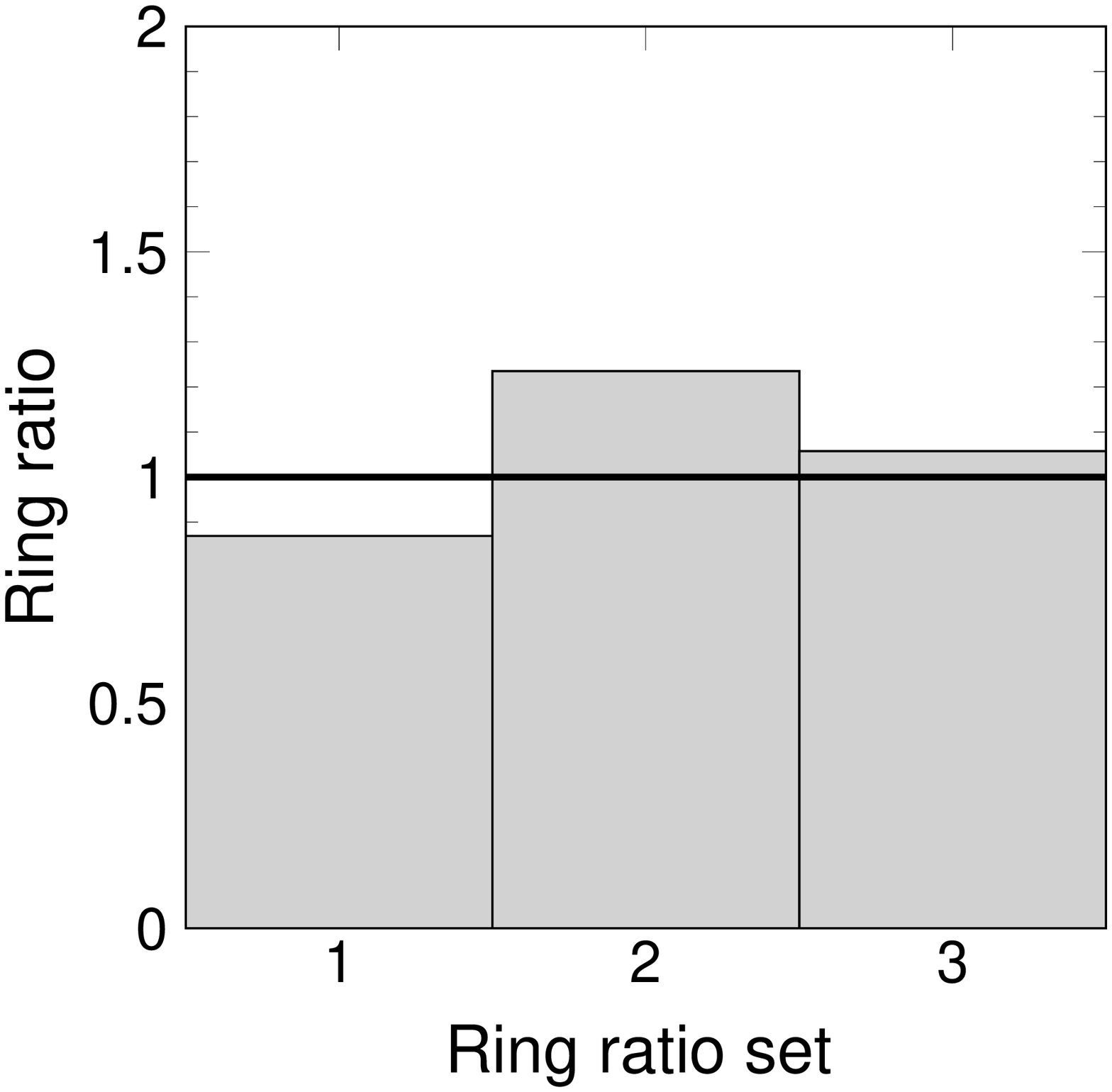}
\caption{Ratio of neutrons detected with different NERO rings for three different runs: production with
$\beta$-delayed neutron emitters (left), background with beam off (center), and background with beam on (right). Histogram bin numbers 1, 2 and 3 correspond to ring ratios $R_{2}/R_{1}$,  $R_{3}/R_{2}$, and $R_{3}/R_{1}$, where $R_{1-3}$ are the innermost, intermediate and external rings. Statistical errors are negligible. 
}
\label{fig:NEROrings}
\end{center}
\end{figure}

Analysis of the ring-counting ratios for background runs (self-trigger mode) and production runs (external trigger mode) support the idea of an external and a beam-induced background source. As shown in Fig.~\ref{fig:NEROrings} (left), measurements performed with NERO in external trigger mode (i.e. from $\beta$ decays in the BCS) with $\beta$-delayed neutron emitters showed that the neutron counting rates were higher for the innermost ring and systematically decreased for the outer rings, in agreement with the results shown in Fig.~\ref{fig:NEROeffRing} (right). On the other hand, background runs with beam off showed the opposite trend, with high rates in the outer ring, gradually decreasing for the inner ones (Fig.~\ref{fig:NEROrings}, center).  
This result suggests that these runs were mainly affected by an external source of background neutrons, most probably related to cosmic rays. Finally, background runs with beam on target showed an intermediate situation that could be explained as arising from a combination of external and internal sources (Fig.~\ref{fig:NEROrings}, right). Energy spectra obtained for background runs with the
ADCs show the wall-effect shape expected for neutrons
(see lower spectra in Fig.~\ref{fig:ADC}). 
Electronic and $\gamma$-ray contributions
are largely below the discriminator thresholds. 

\section{Measurement of $P_{n}$}
\label{sec:measurement}
The NERO detector, together with the BCS, has been employed in numerous r-process motivated experiments performed at
NSCL~\cite{Mon06,Per09,Hos05,Hos09,Qui09}. The exotic nuclei of interest are implanted in a 40$\times$40-pixel DSSD in the BCS. $\beta$-decays are also detected in the DSSD and can be 
position-correlated to previously implanted ions during a maximum correlation rime $t_{c}$.
$P_{n}$ values were determined using the number of neutrons $N_{\beta  n}$ detected in coincidence with an implantation-correlated $\beta$-decay event, 
according to the equation:
\begin{equation}
P_{n}=\frac{N_{\beta  n}-B_{\beta n}-N_{\beta \beta  n}}{\epsilon_{n}  N_{\beta}},
\label{eq:Pn}
\end{equation}
where $B_{\beta n}$ is the number of background coincidences between $\beta$-like events (including real $\beta$ decays and background from the BCS) and neutrons, 
and $N_{\beta}$ is the number of $\beta$-decaying mother nuclei. $N_{\beta \beta  n}$ is the number of detected $\beta$-delayed neutrons from 
the daughter nuclei and needs to be subtracted from
$N_{\beta  n}$. 
For the nuclear species analyzed 
in Refs.~\cite{Mon06,Per09,Hos05,Hos09,Qui09}, $\beta$-neutron coincidences associated with descendant nuclei other than the $\beta$-decay daughter were negligible. In this case, using the Batemann equations~\cite{Cet06}, it is possible to write explicitly the value of $N_{\beta \beta  n}$ as:
\begin{equation}
N_{\beta \beta  n} = (1-P_{n})  C,
\label{eq:Nbbn}
\end{equation}
where $C$ is a constant given by:
\begin{equation}
C = \frac{\lambda_{2} P_{{nn}} N_{\beta} \epsilon_{n}}{\lambda_{2}-\lambda_{1}} \left[
1-e^{{-\lambda_{1}t_{c}}}-\frac{\lambda_{1}}{\lambda_{2}} \left( 1 - e^{{-\lambda_{2} t_{c}}} \right)
\right].
\label{eq:C}
\end{equation}
In this equation, $P_{{nn}}$ is the neutron-emission probability of the daughter nucleus, and $\lambda_{1}$ and $\lambda_{2}$ are the decay constants of the mother and daughter nuclei, respectively. Inserting Eq.~\ref{eq:Nbbn} and Eq.~\ref{eq:C} into Eq.~\ref{eq:Pn}, and rearranging terms:
\begin{equation}
P_{n}=
\frac{N_{\beta  n}-B_{\beta n}-C}{\epsilon_{n} N_{\beta}-C}.
\label{eq:Pnlong}
\end{equation}
The value of $N_{\beta}$ for a given nucleus is calculated as the product of the total number of implantations in the DSSD, and the $\beta$-detection efficiency. 

The NERO background rate given in Sec.~\ref{sec:background} is the  \textquotedblleft free$\textquotedblright$ neutron background rate without any coincidence requirements. In practice, however,  $P_{n}$ values are determined from neutrons measured in coincidence with $\beta$ decays. The number of background $\beta$-neutron coincidences for a given nucleus can be written as 
$B_{\beta n}=B_{n}(\beta)+B_{n}(B_{\beta})$; 
where $B_{n}(\beta)$ is the number of \textquotedblleft free$\textquotedblright$ background neutrons in random coincidence with parent $\beta$ decays, and $B_{n}(B_{\beta})$ is the number of \textquotedblleft free$\textquotedblright$ background neutrons in random coincidence with background $\beta$-like events in the BCS.

%

$B_{n}(B_{\beta})$ as a function of time and detector pixel can be reliably estimated from the  $\beta$-neutron coincidence rates outside of the correlation window of any ion implantations. $B_{n}(B_{\beta})$
for a specific parent nuclide can then be calculated by summing the specific backgrounds at the time and location of each individual ion implantation event.  The high granularity of the DSSD detector greatly reduces this background.

$B_{n}({\beta})$ can be calculated as the product of the number of parent $\beta$ decays detected
$N_{\beta}$ 
and the probability for at least one \textquotedblleft free$\textquotedblright$ background neutron to be detected in random coincidence with each parent $\beta$ decay, i.e.:
\begin{equation}
B_{n}({\beta})=N_{\beta} \sum_{k=1}^{\infty} \frac{(R_{b} \tau)^{k}}{k!} e^{-R_{b} \tau}=N_{\beta}(1-e^{-R_{b} \tau}),
\label{eq:Bnbeta}
\end{equation}
where $\tau$ is the TDC time window defined in Sec.~\ref{sec:moderation}, and  $R_{b}$ is the \textquotedblleft free$\textquotedblright$ neutron background rate. For a typical value $R_{b}\simeq$10~s$^{-1}$, about 0.2\% of the detected parent $\beta$ decays are in random coincidence
with a background neutron, setting the order of magnitude of the lowest $P_{n}$ values that can be measured with NERO under these conditions.

The $P_{n}$ values and their errors obtained in various NSCL experiments  can be found in Refs.~\cite{Mon06,Per09}. The  $P_{n}$ values measured in these experiments 
agree well with perviously measured well established data.

\section{Summary and conclusions}
The neutron detector NERO has been built at 
NSCL enabling the measurement of $\beta$-delayed neutron emission probabilities of r-process nuclei
with fast rare isotope beams. The specific design was motivated by the requirement of achieving a high, energy-independent
neutron detection efficiency up to $\approx$ 1 MeV, and accommodating a large pixelated $\beta$-counting system, necessary to perform measurements with fragmentation beams. MCNP simulations were carried out during the design phase to find the optimum configuration.

Studies of the detector efficiency at various neutron energies were performed with a
$^{252}$Cf source, and with neutrons from a number of
resonant and non-resonant reactions at ISNAP. MCNP calculations reproduce reasonably well the energy dependence of the detector efficiency.
On the other hand, the MCNP calculations slightly overestimate the absolute efficiency of the second and third rings, when the neutrons traverse a larger volume of polyethylene.
An overall scaling of the calculated efficiency 
to the measured data 
can be used to extrapolate the detector efficiency to smaller and larger neutron energies.
The small energy dependence (of about 5\%), for neutron energies below 800~keV,
represents the main contribution from the efficiency correction to the total uncertainty of 
$P_{n}$. 


NERO is currently used with the BCS at NSCL, but can be used with other $\beta$-decay stations
at other rare isotope facilities. 
It will also be used to fully exploit the much higher production rates expected in new generation facilities like FRIB at NSCL, FAIR at GSI and RIBF at RIKEN.

\section*{Acknowledgments}
This work was supported in part by the Joint Institute for Nuclear Astrophysics (JINA) under NSF Grant PHY-02-16783 and the National Superconducting Cyclotron Laboratory (NSCL) under NSF Grant PHY-01-10253.





\end{document}